\documentclass[12pt,letterpaper]{article}
\usepackage{amsmath,amssymb,array,calc,rotating,epsfig,psfrag,amscd, cite}

\makeatletter
\renewcommand\section{\@startsection {section}{1}{\z@}%
                                   {-3.5ex \@plus -1ex \@minus -.2ex}%nn
                                   {2.3ex \@plus.2ex}%
                                   {\normalfont\large\bfseries}}
\renewcommand\subsection{\@startsection{subsection}{2}{\z@}%
                                     {-3.25ex\@plus -1ex \@minus -.2ex}%
                                     {1.5ex \@plus .2ex}%
                                     {\normalfont\bfseries}}
\makeatother

\let\non\nonumber

\let\s=\sigma

\newcommand{\bea}{\begin{eqnarray}}
\newcommand{\eea}{\end{eqnarray}}
\newcommand{\be}{\begin{equation}}
\newcommand{\ee}{\end{equation}}

\newcommand{\p}{\partial}

\newcommand{\C}[1]{$(\ref{#1})$}

\def\IZ{\relax\ifmmode\mathchoice
{\hbox{\cmss Z\kern-.4em Z}}{\hbox{\cmss Z\kern-.4em Z}}
{\lower.9pt\hbox{\cmsss Z\kern-.4em Z}} {\lower1.2pt\hbox{\cmsss
Z\kern-.4em Z}}\else{\cmss Z\kern-.4em Z}\fi}
\def\IR{\relax{\rm I\kern-.18em R}}

\def\one{{\hbox{ 1\kern-.8mm l}}}

\newlength{\bredde}
\def\slash#1{\settowidth{\bredde}{$#1$}\ifmmode\,\raisebox{.15ex}{/}
\hspace*{-\bredde} #1\else$\,\raisebox{.15ex}{/}\hspace*{-\bredde}
#1$\fi}

\newsavebox{\zzzbar}
\sbox{\zzzbar}
  {\setlength{\unitlength}{0.9em}
  \begin{picture}(0.6,0.7)
  \thinlines
  \put(0,0){\line(1,0){0.6}}
  \put(0,0.75){\line(1,0){0.575}}
  \multiput(0,0)(0.0125,0.025){30}{\rule{0.3pt}{0.3pt}}
  \multiput(0.2,0)(0.0125,0.025){30}{\rule{0.3pt}{0.3pt}}
  \put(0,0.75){\line(0,-1){0.15}}
  \put(0.015,0.75){\line(0,-1){0.1}}
  \put(0.03,0.75){\line(0,-1){0.075}}
  \put(0.045,0.75){\line(0,-1){0.05}}
  \put(0.05,0.75){\line(0,-1){0.025}}
  \put(0.6,0){\line(0,1){0.15}}
  \put(0.585,0){\line(0,1){0.1}}
  \put(0.57,0){\line(0,1){0.075}}
  \put(0.555,0){\line(0,1){0.05}}
  \put(0.55,0){\line(0,1){0.025}}
  \end{picture}}

\newcommand{\ena}{\end{eqnarray}}
\newcommand{\beqa}{\begin{eqnarray}}
\newcommand{\eeqa}{\end{eqnarray}}

\newcommand{\g}{\gamma}

\def\g{\gamma}

\def\s{\sigma}

\begin{document}
\begin{titlepage}

\begin{center}

%\hfill \today
%\hfill         \phantom{xxx}         

%\hfill HRI

\vskip 2 cm
{\Large \bf The structure of the $\mathcal{R}^8$ term in type IIB string theory}\\
\vskip 1.25 cm { Anirban Basu\footnote{email address:
    anirbanbasu@hri.res.in} } \\
{\vskip 0.5cm Harish--Chandra Research Institute, Chhatnag Road, Jhusi,
Allahabad 211019, India\\}

\end{center}

\vskip 2 cm

\begin{abstract}
\baselineskip=18pt

Based on the structure of the on--shell linearized superspace of type IIB supergravity,
we argue that there is a non--BPS 16 derivative interaction in the effective action of type IIB string theory of the form $(t_8 t_8 R^4)^2$, which we call the $\mathcal{R}^8$ interaction. It lies in the same supermultiplet as the $G^8 \mathcal{R}^4$ interaction. Using the KLT relation, we analyse the structure of the tree level eight graviton scattering amplitude in the type IIB theory, which leads to the $\mathcal{R}^8$ interaction at the linearized level. This involves an analysis of color ordered multi--gluon disc amplitudes in the type I theory, which shows an intricate pole structure and transcendentality 
consistent with various other interactions. Considerations of S--duality show that the $\mathcal{R}^8$ interaction receives non--analytic contributions in the string coupling at one and two loops.  
Apart from receiving perturbative contributions, we show that the $\mathcal{R}^8$ interaction receives a non--vanishing contribution in the one D--instanton--anti--instanton background at leading order in the weak coupling expansion.  

\end{abstract}

\end{titlepage}

%\pagestyle{plain}
%\baselineskip=18pt
% Try a wider skip
%\baselineskip=19pt
%%%%%%%%%%%%%%%%%%%%%%%%%%%%%%%%%%%%%%%%%%%%%%%%%%%%%%%%%%%%%%%%%%%%%%%%%%%%%%

\section{Introduction}

The structure of the effective action of string theory in various backgrounds contains non--trivial information about the perturbative structure of string amplitudes as well as the non--perturbative duality symmetries of the theory. Though in general the effective action is difficult to calculate, for the maximally supersymmetric theories, some terms can be determined exactly. This is done using information about world sheet loop amplitudes and spacetime supersymmetry, and the interplay of U--duality. The low energy effective action admits a perturbative expansion in $\alpha'$, the inverse string tension.

Some terms in the effective action of type IIB superstring theory in flat 10 dimensional space time have been studied in this context~\cite{Green:1997tv,Green:1997di,Green:1997as,Green:1998by,Green:1999pu,Green:2005ba,Basu:2008cf}. At least at low orders in the momentum (and hence $\alpha'$) expansion, they are all BPS interactions in the low energy effective action. These interactions are of the form $\mathcal{R}^4, D^4 \mathcal{R}^4$ and $D^6 \mathcal{R}^4$ and several other interactions related to them by supersymmetry. Type IIB string theory is conjectured to have an exact $SL(2,\mathbb{Z})$ symmetry in 10 dimensions. The coefficients of these purely gravitational terns in the effective action, in the Einstein frame, are $SL(2,\mathbb{Z})$ invariant modular forms on the fundamental domain of $SL(2,\mathbb{Z})$. Several other interactions related to these by supersymmetry have coefficients which are $SL(2,\mathbb{Z})$ covariant modular forms depending on the field content of the interactions. Needless to say, these interactions are special because they are BPS. 

 It is interesting to also look at non--BPS interactions in the effective action. Though they do not satisfy simple non--renormalization theorems like their BPS counterparts and hence are much more difficult to determine, they provide valuable information about the structure of the theory. Also such operators, unlike the BPS ones, are generic. In this work, we shall analyze a simple non--BPS interaction in the effective action of type IIB string theory in 10 flat dimensions. This is a purely gravitational interaction of the form $\mathcal{R}^8$. Our aim is to initiate a study of its simplest properties based on supersymmetry, S--duality and the properties of type IIB string theory. 

In section 2, we begin by deriving the spacetime structure of the $\mathcal{R}^8$ interaction based on considerations of on--shell linearized supersymmetry. We also discuss other interactions which lie in the same supermultiplet as the $\mathcal{R}^8$ term. In particular, this involves a 32 fermion interaction with a simple spacetime structure. In section 3, we then describe how the structure of the $\mathcal{R}^8$ interaction arises from a tree level string amplitude. This is done using the results of~\cite{Kawai:1985xq}, and hence involves a study of the properties of the color ordered 8 gluon amplitude in type I string theory at the tree level. This necessarily involves an analysis of multi--gluon disc amplitudes, which is considered in the various appendices. Since the pole structure of the 8 graviton amplitude involves the knowledge of various terms in the effective action at lower orders in the $\alpha'$ expansion, this also leads us to discuss the structure of the 4, 5, 6 and 7 gluon disc amplitudes to the desired order in the $\alpha'$ expansion in the appendices. Based on the general structure of supersymmetry, this leads to multi--graviton amplitudes. We next briefly discuss the leading (in the $g_s$ expansion) non--perturbative contribution the $\mathcal{R}^8$ interaction receives, based on its spacetime structure and the constraints imposed by the fermionic zero modes in the D--instanton background. Section 4 analyses very schematically the constraints imposed by supersymmetry and S--duality on the $\mathcal{R}^8$ interaction. Though this discussion is rather qualitative, simple arguments show that there is a non--analytic (in $g_s$) one and two loop contribution to the $\mathcal{R}^8$ interaction, the nature of which is determined by a certain non--analytic source term in the 4 graviton amplitude which vanishes on--shell, as well as the $\mathcal{R}^4$ contact interaction.   

\section{The type IIB superaction and the $\mathcal{R}^8$ interaction}             
\label{theterm}

Our aim is to deduce the structure of the $\mathcal{R}^8$ interaction, and we begin by stating some relevant facts about the type IIB theory. The fields of type IIB supergravity in 10 dimensions are
\be e_\mu^{~a}, \quad B_{\mu\nu}^M, \quad V_{\pm}^{M}, \quad C_{\mu\nu\lambda\rho}, \quad \psi_\mu , \quad \lambda. \ee
Among the bosonic fields $e_\mu^{~a}$ is the vielbein, and $B_{\mu\nu}^M$ is the $SL(2,\mathbb{R})$ doublet which contains the NS--NS and R--R 2 form potentials for $M=1$ and $2$ respectively. $V_{\pm}^M$ contains the complex scalar that parametrizes the coset space $U(1) \backslash SL(2,\mathbb{R})$, where $M=1,2$ and $\pm$ is the $U(1)$ charge $\pm 1$. Also $C_{\mu\nu\lambda\rho}$ is the R--R 4 form potential with self--dual 5 form field strength.   

The fermionic fields include the complex gravitino $\psi_\mu$ which has $U(1)$ charge $3/2$, and the complex dilatino $\lambda$ which has $U(1)$ charge $1/2$. $\psi_\mu$ and $\lambda$ are chiral fermions which have opposite chirality. They are 16 component fermions.  

The field strengths of the various form fields are~\cite{Schwarz:1983qr}
\bea F_{\mu\nu\rho}^M &=& 3 \p_{[\mu} B^M_{\nu\rho]}, \non \\
F_{\mu\nu\rho\lambda\s} &=& 5 \p_{[\mu} C_{\nu\rho\lambda\s]} + \frac{5}{8} i \epsilon_{MN} B^M_{[\mu\nu} F^N_{\rho\lambda\s]}.\ \eea

The 3 form field strengths are combined into the $SL(2,\mathbb{R})$ invariant combinations
\be G_{\mu\nu\rho} = -\epsilon_{MN} V^M_+ F^N_{\mu\nu\rho}, \quad G^*_{\mu\nu\rho} = -\epsilon_{MN} V^M_- F^N_{\mu\nu\rho}, \ee  
which carry $U(1)$ charges 1 and $-1$ respectively. 

One can gauge fix the elements of the coset space $U(1) \backslash SL(2,\mathbb{R})$ to obtain the physical degrees of freedom. This is done by choosing
\be \label{gauge} \left( \begin{array}{cc} V^1_+  & V^2_+ \\ V^1_- & V^2_- \end{array}\right) = \frac{1}{\sqrt{2\tau_2}}\left( \begin{array}{cc} 1 & \tau \\ 1 & \bar\tau \end{array}\right), \ee
where $\tau = C_0 + i e^{-\phi}$, where $C_0$ is the R--R pseudoscalar and $\phi$ is the dilaton. In order to preserve the choice of gauge \C{gauge}, one has to add to the supersymmetry transformation of each field a $U(1)$ violating term
\be \label{add} \delta \Phi =  \frac{iq}{2} (\bar\epsilon \lambda^*  - \bar\epsilon^* \lambda) \Phi,\ee
where $q$ is the $U(1)$ charge of $\Phi$, and $\epsilon$ is the complex supersymmetry fermionic parameter with $U(1)$ charge 1/2. 

The gauge fixed supersymmetry transformations for the various fields involve both $\epsilon$ and $\epsilon^*$. To begin constructing the superaction, we consider only some of the terms involving $\epsilon$ that conserve $U(1)$, for some of the fields. The relevant supersymmetry transformations are given by \C{susy}.
The $U(1)$ violating terms are simply given by adding \C{add} to \C{susy}.

\subsection{The non--BPS superaction}

We now construct a superaction starting from the on--shell linearized superspace for type IIB supergravity~\cite{Howe:1983sra}. Consider the chiral superfield $\Phi(y,\theta)$ where 
\be \label{thetaexp}y^\mu = x^\mu - i (\bar\theta \g^\mu \theta), \ee
which satisfies
\be \label{constraints}\bar{D} \Phi = 0, \quad D^4 \Phi = \bar{D}^4 \bar\Phi , \ee  
where in the $(y,\theta)$ coordinate system
\bea D_\alpha = \frac{\p}{\p \theta^\alpha} + 2i (\g^\mu \theta^*)_\alpha \p_\mu, \quad \bar{D}_\alpha = -\frac{\p}{\p \theta^{*\alpha}} .\eea 
The superfield $\Phi$ can be expanded in powers of $\theta$ where the fields at a fixed order in the $\theta$ expansion are functions of $y$. While the first constraint in \C{constraints} enforces chirality as discussed above, the second constraint imposes the field equations of supergravity as well as the self--duality of the 5 form field strength. Using \C{susy}, and keeping only the $U(1)$ charge conserving bosonic fields in the $\theta$ expansion at even powers of $\theta$ which is good enough for our purposes, we see that the superfield $\Phi$ admits the expansion (dropping various irrelevant numerical factors)
\bea \label{expand} \Phi(y,\theta) &=& \tau + \tau_2 (\bar\theta^* \lambda) + \tau_2 (\bar\theta^* \g^{\mu\nu\rho} \theta) \hat{G}_{\mu\nu\rho} + \tau_2 (\bar\theta^* \g^{\mu\nu\lambda} \theta) (\bar\theta^* \g_\mu \p_\nu \psi_\lambda) \non \\ && + \tau_2 (\bar\theta^* \g^{\mu\nu\lambda} \theta) (\bar\theta^* \g_\mu^{~\rho\s} \theta) R_{\nu\lambda\rho\s} +\tau_2 (\bar\theta^* \g^{\mu\nu\lambda} \theta) (\bar\theta^* \g_\mu \g^{\rho_1 \ldots \rho_5} \g_\nu \theta) \p_\lambda F_{\rho_1 \ldots \rho_5} \non \\ &&+ O(\theta^5).\eea
There are many $U(1)$ violating terms in \C{expand} which arise from gauge fixing the supersymmetry transformations, for example, at $O(\theta^2)$ there is a $(\bar\theta^* \lambda)^2$ term. Also, what is actually obtained is the Riemann curvature term at the linearized level ($R = d \omega +\ldots$) and we have written its non--linear completion.   

One can construct a $1/2$ BPS superaction by integrating an arbitrary function of the chiral superfield $\Phi$ over the Lorentz invariant chiral half of superspace involving only $\theta$. Thus, these are F terms. This leads to, among other terms, the $\mathcal{R}^4$, $\psi^* \lambda^{15}$ and $\lambda^{16}$ interactions~\cite{Green:1998by,Green:1999qt}.   

We want to construct a class of non--BPS interactions in the type IIB effective action using the superfield \C{expand}. These are given by integrating an arbitrary function of $\Phi$ over the whole of superspace, which can be done in a Lorentz invariant way. Thus, this is a D term. We write the superaction as
\be \label{saction} \int d^{16} \theta d^{16} \theta^*  f (\Phi, \bar{\Phi}),\ee
where we define
\be d^{16} \theta = \frac{1}{16 !} \epsilon^{\alpha_1 \ldots \alpha_{16}} d\theta_{\alpha_1} \cdots d\theta_{\alpha_{16}}, \ee
and similarly for $d^{16} \theta^*$, which yields non--BPS interactions in the type IIB effective action. In particular, it yields a purely gravitational term
\be \label{R8}
\frac{1}{(4!)^2} g (\tau, \bar\tau) \int d^{16} \theta (\theta^4 R)^4 \int d^{16} \theta^* (\theta^* R)^4\ee
where
\be \label{lin1} g(\tau,\bar\tau)  =  \Big( \tau_2 \frac{\p}{\p\tau}\Big)^4 \Big( \tau_2\frac{\p}{\p \bar\tau}\Big)^4 f(\tau,\bar\tau),\ee
and
\be \theta^4 R \equiv (\bar\theta^* \g^{\mu\nu\lambda} \theta) (\bar\theta^* \g_\mu^{~\rho\s} \theta) R_{\nu\lambda\rho\s}, \quad \theta^{*4} R \equiv (\bar\theta \g^{\mu\nu\lambda} \theta^*) (\bar\theta \g_\mu^{~\rho\s} \theta^*) R_{\nu\lambda\rho\s}.\ee 
At the linearized level, the ${\mathcal{R}}^8$ interaction in \C{R8} must yield the 8 graviton amplitude in type IIB string theory. This multi--graviton amplitude interaction is non--BPS and should not satisfy the strong non--renormalization theorems that the BPS interactions like $\mathcal{R}^4, D^4 \mathcal{R}^4$ and $D^6 \mathcal{R}^4$ satisfy, which are expressed as solutions to a simple Poisson equation on moduli space, and receive only a finite number of perturbative contributions.    

The $\mathcal{R}^8$ interaction has the same number of derivatives as the $D^8 \mathcal{R}^4$ interaction, which is a non--BPS interaction as well. We expect these two interactions to be a part of the same non--linear supermultiplet because of maximal supersymmetry. The moduli dependent coefficients of these interactions in the Einstein frame are $SL(2,\mathbb{Z})$ invariant modular forms. Because of maximal supersymmetry, these coefficients should be the same, which we shall assume to be true, which will be found to be self--consistent in our analysis with the structure of various other interactions. We shall provide some evidence for this at the level of the tree amplitude, which involves an analysis of the structure of the 8 gluon disc amplitude in the type I theory. While we do not extract the final coefficient, the structure that follows from the multi--gluon amplitudes provides strong evidence for this, as it involves many integrals all of which yield Riemann zeta functions only of a fixed transcendentality. On the side, our analysis gives non--trivial information about multi--gluon tree level scattering amplitudes in the type I theory. These amplitudes lead to various higher derivative pure Yang--Mills interactions, which will lead to relations between various such interactions based on supersymmetry.
We shall see that all these various observations fit in well with some existing knowledge of higher derivative corrections.

The fact that $\mathcal{R}^8$ and $D^8\mathcal{R}^4$ interactions should lie in the same supermultiplet does not follow from our linearized analysis. 
To see this, apart from \C{susy}, also consider the sypersymmetry transformations \C{susymore} to construct the superfield $\Phi$ at higher powers in $\theta$. 
Thus in \C{expand}, at $O(\theta^8)$, there is a term of the form $\p^4 \bar\tau$, and similarly in $\bar\Phi$ which has a term of the form $\p^4 \tau$. This leads to an interaction of the form $(\p^4 \tau)(\p^4 \bar\tau) R^4$. However given the structure of the $R$ and $\p^4 \bar\tau$ terms in $\Phi$, and the $R$ and $\p^4 \tau$ terms in $\bar\Phi$, and the structure of the Grassmann integrals in \C{saction}, it follows that we do not get a term of the form $D^8 \mathcal{R}^4$ on integrating by parts, which has the required spacetime structure, which is fixed by looking at the 8 graviton tree amplitude, for example. Thus all this gives us is an interaction with a non--constant $\tau$, not the one we want. Another possibility would be to use \C{thetaexp} and obtain the product of $(\theta^4 )^2 (\theta^{*2} \theta^2 \p^2 R)^2$ and $(\theta^{*4} )^2 (\theta^{*2} \theta^2 \p^2 R)^2$ in the expansion of $\Phi$. However, from the index contractions it follows that it does not yield the required term. 

The precise spacetime structure of the non--BPS $\mathcal{R}^8$ interaction follows from \C{R8}. The integral over the 16 Grassmann parameters $\theta$ yield the standard $\mathcal{R}^4$  interaction given by $t_8 t_8 R^4$, and so we get that
\be \label{structure} \mathcal{R}^8 = (t_8 t_8 R^4 )^2.\ee
Thus our linearized superspace analysis shows that the spacetime structure of the $\mathcal{R}^8$ interaction is given by the square of the spacetime structure of the $\mathcal{R}^4$ interaction. In general multi--graviton amplitudes have a complicated structure and obtaining the precise space--time structure of the gravitational interactions at the full non--linear level that follow from it is quite difficult. However, for the $\mathcal{R}^8$ interaction, linearized superspace gives the spacetime structure in \C{structure} very easily. 

\subsection{Other non--BPS interactions in the same supermultiplet}

Let us consider some other non--BPS interactions in the same supermultiplet as the $\mathcal{R}^8$ interaction. Consider the $G^8 \mathcal{R}^4$ interaction which also follows from \C{saction} on using \C{expand}. This is given by
\be \label{G8R4}
\frac{1}{4! 8!} h (\tau, \bar\tau) \int d^{16} \theta (\theta^2 \hat{G})^8 \int d^{16} \theta^* (\theta^* R)^4\ee
on keeping only the bosons in $\hat{G}$, where
\be \label{lin2} h(\tau,\bar\tau)  =  \Big( \tau_2 \frac{\p}{\p\tau}\Big)^8 \Big( \tau_2\frac{\p}{\p \bar\tau}\Big)^4 f(\tau,\bar\tau),\ee
and
\be \theta^2 \hat{G} \equiv (\bar\theta^* \g^{\mu\nu\rho} \theta)  \hat{G}_{\mu\nu\rho} .\ee 
This is the $SL(2,\mathbb{Z})$ covariant generalization of the $SL(2,\mathbb{Z})$ invariant $(G G^*)^{4} \mathcal{R}^4$ interaction~\cite{Berkovits:1998ex}, which should be a part of the same supermultiplet. Our analysis indeed supports this claim. 

From \C{expand} and \C{saction}, we also obtain the maximally fermionic interaction
\be \label{int3} \frac{1}{(16 !)^2} q (\tau,\bar\tau) \int d^{16} \theta (\bar\theta^* \lambda)^{16} \int d^{16} \theta^* (\bar\theta \lambda^*)^{16},\ee
where
\be \label{lin3} q(\tau,\bar\tau) =  \Big( \tau_2 \frac{\p}{\p\tau}\Big)^{16} \Big( \tau_2\frac{\p}{\p \bar\tau}\Big)^{16} f(\tau,\bar\tau).\ee

We shall discuss the possible relation between $g(\tau,\bar\tau)$, $h(\tau,\bar\tau)$ and $q(\tau,\bar\tau)$ at the non--linear level, based on the constraints imposed by S--duality and supersymmetry in section 4.

\section{The structure of the $\mathcal{R}^8$ interaction in type IIB string theory}

Our aim is to obtain the nature of the $\mathcal{R}^8$ interaction in type IIB string theory. Though this multi--graviton amplitude has a complicated spacetime structure, \C{structure} gives us the full non--linear completion of the interaction.

\subsection{The structure from a tree level amplitude calculation}

We consider the calculation of the 8 graviton amplitude at tree level, in the RNS formalism, in appendix D. Dropping overall factors, this yields a contact term in the effective action given by
\be \zeta (7) \int d^{10} x \sqrt{-g} e^{-2\phi} \mathcal{R}^8\ee
in the string frame, as analyzed in \C{8ptint}. Thus by supersymmetry, the tree level coefficient of the $G^8 \mathcal{R}^4$ interaction defined by \C{G8R4} must also be proportional to $\zeta (7)$. This interaction should be in the same multiplet as the $(G G^*)^4\mathcal{R}^4$ interaction which indeed has this coefficient~\cite{Berkovits:1998ex}.

\subsection{The leading non--perturbative contribution}

Apart from receiving perturbative contributions, the $\mathcal{R}^8$ must receive non--perturbative contributions to be consistent with S--duality. These are given by calculating the contribution this term receives in the D--(anti)--instanton background, which is determined by the structure of the zero modes in the instanton background. Given the spacetime structure of the $\mathcal{R}^8$ term \C{structure}, it is easy to see what the contribution is at leading order in the string coupling. One factor of $t_8 t_8 R^4$ obtained by integrating over $\theta$ saturates the 16 zero modes in the instanton background exactly as in the case of the $\mathcal{R}^4$ interaction, while the other factor of $t_8 t_8 R^4$ obtained by integrating over $\theta^*$ saturates the 16 zero modes in the anti--instanton background~\cite{Green:1997tv,Green:1997di,Green:1997me}. Thus the leading non--perturbative contribution the $\mathcal{R}^8$ term receives is of the form
\be \tau_2^n e^{2\pi i (\tau -\bar\tau)} = \tau_2^n e^{-4\pi e^{-\phi}},\ee   
where $n$ is determined by the tree level contribution in the instanton background. This also follows easily by noting that the $\lambda^{16} \lambda^{*16}$ interaction is in the same supermultiplet, and $\lambda$ ($\lambda^*$) soaks up 1 zero mode in the instanton (anti--instanton) background. 

Such non--extremal instanton contributions have been analyzed in~\cite{Bergshoeff:2004fq,Bergshoeff:2005zf}, where it was suggested that they contribute to the $\mathcal{R}^8$ coupling. We see the structure arising directly using the superspace action, and the zero mode analysis. 

\section{Relations among the moduli dependent couplings of the various non--BPS interactions}
\label{susyrel}

Among the many interactions that follow from \C{saction}, the 
\be \label{3int} \mathcal{R}^8 , \quad G^8 \mathcal{R}^4, \quad \lambda^{16} \lambda^{*16} \ee
 interactions have coefficients $g(\tau,\bar\tau)$, $h(\tau,\bar\tau)$ and $q(\tau,\bar\tau)$ respectively,  which follow from the terms in the action given by \C{R8}, \C{G8R4} and \C{int3}. Thus, from \C{lin1}, \C{lin2} and \C{lin3} if follows that at the linearized level, 
they are related by
\be\label{lin}
 h = \Big( \tau_2 \frac{\p}{\p\tau} \Big)^4 g, \quad q = \Big( \tau_2 \frac{\p}{\p \tau}\Big)^{12} \Big(\tau_2 \frac{\p}{\p \bar\tau}\Big)^{12} g. \ee
At the linearized level, note that $\tau_2$ in \C{lin1}, \C{lin2} and \C{lin3} is a background field that does not fluctuate.

What is the relation among these couplings at the non--linear level? The basic structure of the relation simply follows from S--duality and supersymmetry, as we now explain. Note that all the interactions in \C{3int} and all those related to it by supersymmetry are expressed in the Einstein frame\footnote{In this section, the various interactions are all written in the Einstein frame. In general, it wil be clear from the context which frame we are in.}.  

To get this from the string calculation, we need to convert all the purely gravitational interactions from the string frame to the Einstein frame using
\be g_{\mu\nu}^\s = e^{\phi/2} g_{\mu\nu}^E,\ee   
where $g_{\mu\nu}^\s$ and $g_{\mu\nu}^E$ are the string and Einstein frame metrics respectively. 
In the Einstein frame, the metric is S--duality invariant, and thus the coefficients of all purely gravitational interactions must be $SL(2,\mathbb{Z})$ invariant modular forms. For example, the $\mathcal{R}^8$ interaction yields a term in the effective action in the Einstein frame given by
\be \int d^{10} x \Big(\zeta (7)\tau_2^{7/2} + \cdots \Big)\sqrt{-g}\mathcal{R}^8\ee  
where $\zeta (7)$ is the tree level contribution to the 8 graviton amplitude. All the other interactions have to be converted to the Einstein frame after appropriately defining the various fields in the Einstein frame. From supersymmetry it follows that all the interactions in \C{3int} and all those related to it by supersymmetry have a tree level contribution of the form $\zeta (7)\tau_2^{7/2}$. Naturally all the higher loop (suppressed by factors of $\tau_2^{-2}$) and non--perturbative contributions (suppressed by factors of $e^{2\pi i \tau}$ and $e^{-2\pi i \bar\tau}$) must also have the same structure as the $\mathcal{R}^8$ coupling. 

While the $\mathcal{R}^8$ and $\lambda^{16} \lambda^{*16}$ terms are $SL(2,\mathbb{Z})$ invariant, the $G^8 \mathcal{R}^4$ term is not. The $G^8 \mathcal{R}^4$ term has $SL(2,\mathbb{Z})$ weight $(-4,4)$ and thus its coefficient is a modular form of weight $(4,-4)$, so the whole interaction is S--duality invariant (see \C{modform}). We thus find it convenient to express these interactions as
\be \int d^{10} x \sqrt{-g}\Big( g^{(0,0)}(\tau,\bar\tau)\mathcal{R}^8 + h^{(4,-4)} (\tau,\bar\tau)G^8 \mathcal{R}^4 + q^{(0,0)} (\tau,\bar\tau)\lambda^{16} \lambda^{*16}\Big)\ee 
in the effective action.

To understand the relationship between the various couplings in the $\mathcal{R}^8$ supermultiplet, we use the invariance of the action under supersymmetry transformations
\be \label{inv}\delta S =0,\ee
upto total derivatives. 
We expand the action and the supersymmetry transformations in powers of $\alpha'$ as
\be \label{expa} S = S^{(0)} + \sum_{n=3}^\infty S^{(n)}, \quad \delta = \delta^{(0)} + \sum_{n=3}^\infty \delta^{(n)}, 
\ee
where the terms of order $n$ in the expansion are $O(\alpha'^n)$ suppressed compared to the supergravity action $S^{(0)}$ and the supergravity supersymmetry transformations $\delta^{(0)}$. The terms at $n=1,2$ vanish in the action in \C{expa} because the $\mathcal{R}^2$ and $\mathcal{R}^3$ interactions vanish in the effective action due to supersymmetry, and thus the whole supermultiplets vanish. Thus the corresponding terms also vanish in the expression for $\delta$ in \C{expa}. The $\mathcal{R}^8$ term is part of $S^{(7)}$ in our convention. Thus at $O(\alpha'^7)$ compared to supergravity, \C{inv} leads to
\be \label{noether} \delta^{(0)} S^{(7)} + \delta^{(7)} S^{(0)} + \delta^{(3)} S^{(4)} + \delta^{(4)} S^{(3)} =0. \ee    
From \C{noether} one can judiciously choose maximally fermionic terms in the effective action ~\cite{Green:1998by,Basu:2008cf} and construct corrected supersymmetry transformations such that \C{noether} is satisfied. We shall refer to the last two contributions in \C{noether} as source terms, since they involve $S^{(m)}$ and $\delta^{(n)}$ for $m,n<7$. We shall see their precise role shortly below. This technique has proved powerful in determining several maximally fermionic couplings in the effective action.  

In the absence of a detailed understanding of the various supermultiplets involved, we shall simply very schematically deduce the structure of the relationship between the various couplings in the $\mathcal{R}^8$ multiplet without going into the detailed structure, or keeping track of numerical factors. To start with it is very convenient to consider the 32 fermion terms in $S^{(7)}$ given by
\be S^{(7)} = \int d^{10} x \sqrt{-g} \Big( q^{(0,0)} \lambda^{16} \lambda^{*16} + q^{(1,-1)} \lambda^{16} \lambda^{*15} \psi \Big) +\ldots\ee
where the second interaction comes from $\hat{G}^* \lambda^{*14}\lambda^{16}$ which follows from \C{saction}, and using \C{hat}. Now acting with $\delta^{(0)}$, we get that
\be \label{eqnone}
\delta^{(0)} S^{(7)} = \int d^{10} x \sqrt{-g} \Big( \bar{D}_{-1} q^{(1,-1)} + q^{(0,0)}\Big) \epsilon^*\lambda^{16} \psi \lambda^{*16} + \ldots,\ee
where the modular covariant derivative is given by \C{covder}. Now \C{eqnone} is the only contribution of this type in $\delta^{(0)} S^{(7)}$, and a similar contribution coming from $\delta^{(7)} S^{(0)}$ should preserve the structure of this equation along the lines of~\cite{Green:1998by,Basu:2008cf}. This is because a supervariation of the type
\be \label{eqntwo}\delta^{(7)} \lambda = q^{(0,0)} \epsilon^* \lambda^{15} \lambda^{*14} \psi, \quad \delta^{(7)} \lambda^* = q^{(0,0)}\epsilon^* \lambda^{14} \psi \lambda^{*15} \ee   
acting on the $\lambda^2 \lambda^{*2}$ interaction in $S^{(0)}$ produces precisely the spacetime structure needed. The fact that the coefficient in \C{eqntwo} is $q^{(0,0)}$ must follow from the on--shell closure of the supersymmetry algebra. 
Thus we get an equation of the form
\be  \bar{D}_{-1} q^{(1,-1)} = q^{(0,0)} + \ldots,\ee
where the missing terms of vanishing modular weight come from the $ \delta^{(3)} S^{(4)}$ and $\delta^{(4)} S^{(3)}$ contributions. Among the many terms which do contribute, let us focus on two very simple kinds of terms which contribute to illustrate the structure that arises: (i) the $E_{3/2}$ contribution which comes from $S^{(3)}$~\cite{Green:1998by,Green:1997tv,Green:1997di,Green:1997as}, and the (ii) the $Y$ contribution which comes from $S^{(4)}$. The corrected supervariations $\delta^{(3)}$ ($\delta^{(4)}$) must be proportional to the modular forms coming from $S^{(4)}$ ($S^{(3)}$). 

In the above discussion, $E_{3/2}$ is the coefficient of the $\mathcal{R}^4$ interaction, and receives perturbative contributions only at tree level and one loop given by
\be E_{3/2} = 2 \zeta (3) \tau_2^{3/2} + 4 \zeta (2) \tau_2^{-1/2} + \ldots. \ee
Now in $S^{(4)}$, the $\mathcal{R}^5$ interaction vanishes (see appendix E for details), while the $D^2 \mathcal{R}^4$ interaction also vanishes on--shell. So what is this possible $Y$ contribution?

To see the origin of $Y$~\cite{Green:2006gt,Green:2008uj}, note that the one loop 4 graviton amplitude has a non--analytic piece in the external momenta of the form $s {\rm ln} (-\alpha' s) \mathcal{R}^4$ in the string frame (symmetrized in $s,t$ and $u$), which produces an interaction of the form $({\rm ln} \tau_2) (s+t+u)\mathcal{R}^4$ in the Einstein frame in $S^{(4)}$, which vanishes on--shell. We call this coupling $Y$, thus it receives only a one--loop contribution 
\be Y = \zeta (2){\rm ln} \tau_2 +\ldots \ee 
and non--perturbative contributions due to S--duality. Though this vanishes in $S^{(4)}$, it can contribute 
in $\delta^{(4)}$, because the term in $S^{(3)}$ given by $E_{3/2}\lambda^8\lambda^{*8}$ produces the structure in \C{eqnone} for
\be \label{eqnthree} \delta^{(4)} \lambda = Y\epsilon^* \lambda^9  \psi \lambda^{*8}, \quad \delta^{(4)} \lambda^* = Y\epsilon^* \lambda^{*9} \psi \lambda^8.\ee
Unlike earlier, closure of the supersymmetry algebra does not determine $Y$ as the corresponding term in $S^{(4)}$ vanishes to start with. But its contribution can be inferred from the structure of the non--analytic term in $S^{(4)}$ in the effective action, and can possibly contribute in \C{eqnthree}.              
    
Thus we get that
\be \label{pat}\bar{D}_{-1} q^{(1,-1)} = q^{(0,0)} + Y E_{3/2} + \ldots, \ee
where we have dropped all other possible source term contributions. The pattern in \C{pat} should continue further to yield
\be \bar{D}_{-2} q^{(2,-2)} = q^{(1,-1)} + (D_0 Y) E_{3/2} + Y (D_0 E_{3/2}) + \ldots,\ee
for the $\hat{G}^{*2} \lambda^{*12} \lambda^{16}$ interaction, and so on. This type of iteration stops at
\be \label{e1}\bar{D}_{-12} q^{(12,-12)} = q^{(11,-11)} + \sum_{m,n;m+n=11}(D^m Y ) (D^n E_{3/2})+ \ldots,\ee  
where $q^{(12,-12)}$ and $q^{(11,-11)}$ are the couplings of the $\lambda^{16} \mathcal{R}^4$ and $\hat{G}^{*2}\lambda^{16} \mathcal{R}^3$ interactions respectively, and $D^n$ stands for $n$ modular covariant derivatives acting as $D_{n-1}\ldots D_1 D_0$. 
In fact, 12 is the hightest modular weight of any interaction in this supermultiplet. 
Also note that it is possible that the coefficient of a given interaction splits into several modular forms, each of which satisfies an equation of the type described above, which we have not analyzed. This is known to happen, for example, for the $D^8 \mathcal{R}^4$ interaction~\cite{Basu:2008cf} and for interactions in the 9 dimensional theory~\cite{Green:2008bf}.  

To continue further along the multiplet, again we expect an equation of the form
\be \label{e2} D_{11} r^{(11,-11)} = q^{(12,-12)} + \sum_{m,n;m+n=12}(D^m Y ) (D^n E_{3/2})+ \ldots,\ee
where $r^{(11,-11)}$ is the coefficient of the $\lambda^{15} \psi^* \mathcal{R}^4$ interaction. If it is equal to $q^{(11,-11)}$, then \C{e1} and \C{e2} imply a Poisson equation for for $q^{(12,-12)}$ and $q^{(11,-11)}$ on the fundamental domain of $SL(2,\mathbb{Z})$, with source terms determined by interactions at lower orders in the momentum expansion. Generalizing this argument, one can continue all the way down to
\be D_0 r^{(0,0)} = r^{(1,-1)} + (D_0 Y ) E_{3/2} + Y (D_0 E_{3/2}) + \ldots,\ee    
where $r^{(0,0)}$ and $r^{(1,-1)}$ are the coefficients of the $\mathcal{R}^8$ and $\hat{G}^2 \mathcal{R}^7$ interactions. 
Though we have not analyzed the above structure in any detail, we see that it is completely constrained by supersymmetry and S--duality. 

Note that the introduction of the $Y$ term in our analysis looks strange. First of all, it arises as the coefficient of a term in the effective action which vanishes on--shell. Also in the source term it arises as the coefficient of the corrected supersymmetry transformations \C{eqnthree}, which is not fixed by closure due to vanishing of the corresponding term in the action. Thus our analysis is really inert to the presence of the $Y$ term. We claim that the presence of this term will show up unambiguously in an off--shell version version of the analysis, if it can be done. We now present some evidence for the existence of this term.    

The differential equation satisfied by the $\mathcal{R}^8$ coupling on the fundamental domain of $SL(2,\mathbb{Z})$ must involve source terms of modular weight 0. Our analysis suggests that one such term is of the form
\be Y E_{3/2} = \zeta (2){\rm ln} \tau_2 \Big( 2\zeta (3) \tau_2^{3/2} + 4 \zeta (2) \tau_2^{-1/2}\Big) + \ldots. \ee 
Thus the $\mathcal{R}^8$ interaction must have a 1 loop and a 2 loop contribution of the form $ \zeta (2)\zeta (3) \tau_2^{3/2}{\rm ln} \tau_2$ and $\zeta (2)^2 \tau_2^{-1/2} {\rm ln}\tau_2$ respectively. Thus we expect the $D^8 \mathcal{R}^4$ term to have these contributions as well. In fact, the analytic part of the 1 loop contribution to $D^8 \mathcal{R}^4$ vanishes, while the non--analytic part indeed has this term~\cite{Green:2008uj}. A direct 1 loop calculation of the 8 graviton amplitude should confirm our argument. We expect the $R^8$ coupling to split into a sum of $SL(2,\mathbb{Z})$ invariant modular forms, each of which satisfies a Poisson equation as discussed above.

\section{Some generalities}

Let us discuss a general point about the structure of the effective action following from the tree level multi--gluon, and multi--graviton amplitudes, which should hold beyond perturbation theory, and in fact, exactly.
We have so far assumed that for the type II string effective action, the various interactions that arise with the same mass dimension are in the same supermultiplet, and hence they have the same couplings\footnote{If the coupling for a specific interaction splits into a sum of different modular forms consistent with S--duality, the same modular forms appear as couplings for all interactions in the supermultiplet.}. We have assumed similarly for the type I effective action for interactions involving pure Yang--Mills couplings involving a single trace. This is a consequence of maximal supersymmetry, as it is expected that all the interactions at a given order in the $\alpha'$ expansion will be in an irreducible representation of the supersymmetry algebra. This follows from the fact that all these fields do form an irreducible representation of the superalgebra. The fields of the type IIB theory transform under an irreducible representation of the $N=2$ gravity algebra, while the fields of the type I theory decompose into the sum of 2 irreducible representations of the $N=1$ algebra: $N=1$ gravity, and $N=1$ Yang--Mills.  For single trace terms involving only Yang--Mills fields in the effective action this amounts to replacing a $D^2$ by $F$, while one has to replace $D^2$ by $R$ for purely gravitational terms in the effective action\footnote{Note that for the type I theory, all pure Yang--Mills interactions with the same number of derivatives no more lie in the same supermultiplet once double trace interactions are introduced. In fact, the single and double trace interactions have very different properties at higher loops~\cite{Berkovits:2009aw}.}. This was evident from our discussion, and also the results of the various integrals, which had a priori no reason to reproduce the same transcendental structure. Thus from the four point tree level calculations, it follows that the coefficients of the various pure Yang--Mills terms in the effective action should only involve Riemann zeta functions, while the purely gravitational terms should involve Riemann zeta functions only of odd transcendentality. This structure then follows for the entire supermultiplet.     

To end with, we now report some more multi--gluon integrals that corroborate this claim, which can be done with very little extra work given the details in appendix E. As we discussed in appendix E, the integral \C{genint} is an integral involving 3 logarithms in the 7 point amplitude, as well as an integral involving 2 logarithms in the 8 point amplitude\footnote {upto an overall sign}. Now each logarithm in an $N$ point gluon amplitude yields a factor of $\alpha' \p^2$, and this precisely accounts for the extra factor of $\alpha' F$ and the difference of 1 logarithm when being equated with the integral for an $N+1$ point function. We now show that at an arbitrary order in the momentum expansion for the 7 and 8 point amplitudes, this equality of the integrals hold for a specific set of integrals, and further show that they all yield Riemann zeta functions. For the 7 and 8 point amplitudes, we have that
\bea &&\int_0^1 dp \int_0^1 dy \int_0^1 dw \int_0^1 dx \int_0^1 dz \frac{1}{pywxz} {\rm ln} (1-pyw) {\rm ln} (1-wxz) ({\rm ln} p)^N \non \\ &=& -\frac{1}{N+1} \int_0^1 dp \int_0^1 dy \int_0^1 dw \int_0^1 dx \frac{1}{pywx} {\rm ln} (1-py) {\rm ln} (1-wxy) ({\rm ln} p)^{N+1} \non \\ 
&=&(-1)^N N! \int_0^1 \frac{dx}{x} Li_3 (x) Li_{N+3} (x) \non \\ &=& (-1)^N N! \Big( \zeta (3) \zeta (N+4) - \zeta (2) \zeta (N+5) + \sum_{m,n=1}^\infty \frac{1}{m^{N+5} n(m+n)} \Big),\eea
and we next use the relation
\be \label{needsum} \sum_{m,n=1}^\infty \frac{1}{m^P n(m+n)} = \sum_{Q=0}^{P-2} \zeta (P -Q, Q+2) + 2\zeta (P+1,1).\ee
In \C{needsum}, for $P$ odd the terms from the two ends of the summation pair up and simplify using \C{stuffle}, while for $P$ even there is an extra term $\zeta (P/2+1,P/2+1)$ which also simplifies using \C{stuffle}. Finally the $2\zeta(P+1,1)$ term simplifies using \C{zetan1}, and the entire answer is expressed only in terms of Riemann zeta functions of total transcendentality $N+7$. This completes the argument. 

\vspace{.5cm}

{\bf{Acknowledgements:}} I am thankful to Ashoke Sen for bringing reference~\cite{Stieberger:2009rr} to my notice, and to D. Surya Ramana for useful comments.

\section{Appendix}

\appendix

\section{Supersymmetry transformations}

We shall need the structure of the supersymmetry transformations for the various fields. We mention only the relevant transformations. The $O(\epsilon)$ terms in the supervariation of $\tau, e_\mu^{~a}, B, \lambda$ and $\psi$ are~\cite{Schwarz:1983qr}

\bea \label{susy}\delta^{(0)} \tau &=& -2 i\tau_2 \bar\epsilon^* \lambda, \non \\ 
\delta e_\mu^{~a} &=& i\bar\epsilon^* \gamma^a \psi_\mu^* + \ldots, \non \\ 
\delta^{(0)} B^M_{\mu\nu} &=&  4i V^M_- \bar\epsilon^* \g_{[\mu} \psi_{\nu]} + \ldots, \non \\ 
\delta^{(0)} \lambda &=& -\frac{i}{24} \g^{\mu\nu\rho} \epsilon \hat{G}_{\mu\nu\rho} +\ldots, \non \\ \delta^{(0)} \psi_\mu  &=& D_\mu \epsilon +\frac{i}{480} \g^{\mu_1 \cdots \mu_5} \g_\mu \epsilon \hat{F}_{\mu_1 \cdots \mu_5} +\ldots,\eea
where the hatted supercovariant fields in \C{susy} are given by 
\bea \label{hat}\hat{G}_{\mu\nu\rho} &=& G_{\mu\nu\rho} - 3 \bar\psi_{[\mu} \g_{\nu\rho]} \lambda - 6i \bar\psi^*_{[\mu} \g_\nu \psi_{\rho]}, \non \\ \hat{F}_{\mu\nu\rho\lambda\s}&=& F_{\mu\nu\rho\lambda\s} -5 \bar\psi_{[\mu} \g_{\nu\rho\lambda} \psi_{\s]} - \frac{1}{16} \bar\lambda \g_{\mu\nu\rho\lambda\s} \lambda. \eea

The $O(\epsilon)$ terms in the supervariation of $C,\psi^*$ and $\lambda^*$ are
\bea \label{susymore}
\delta^{(0)} C_{\mu\nu\lambda\rho} &=& -\bar\epsilon^* \g_{[\mu\nu\lambda} \psi_{\rho]}^* 
%- \frac{3}{8}i \epsilon_{MN} B^M_{[\mu\nu} \delta B^N_{\lambda\rho]} 
+\ldots, \non \\ \delta^{(0)} \psi_\mu^* &=& \frac{1}{96} \Big(\g_\mu^{~\nu\rho\lambda} \hat{G}^*_{\nu\rho\lambda} -9 \g^{\nu\lambda} \hat{G}^*_{\mu\nu\lambda}\Big)\epsilon+ \ldots, \non \\ \delta^{(0)} \lambda^* &=& i \g^\mu \epsilon \frac{\p_\mu \tau^*}{2\tau_2} + \ldots .  \eea

\section{Modular forms of $SL(2,\mathbb{Z})$} 

A modular form $\Psi^{(m,n)} (\tau,\bar\tau)$ of weight $(m,n)$ of $SL(2,\mathbb{Z})$ transforms as
\be \label{modform} \Psi^{(m,n)} (\tau', \bar\tau') = (c\tau + d)^m(c\bar\tau + d)^n \Psi^{(m,n)} (\tau,\bar\tau)\ee
under modular transformations
\be \tau' = \frac{a\tau + b}{c\tau +d},\ee
where $a,b,c,d \in \mathbb{Z}$ and $ad-bc=1$.

In the Einstein frame, a field of type IIB string theory which carries $U(1)$ charge $(q,-q)$ in supergravity, transforms as a  modular form of $SL(2,\mathbb{Z})$ of weight $(-q/2,q/2)$ .

Modular covariant derivatives $D_m$ and $\bar{D}_n$ are defined by~\cite{Green:1998by}
\be \label{covder} D_m = i\Big( \tau_2 \frac{\p}{\p \tau} - \frac{im}{2}\Big), \quad \bar{D}_n = -i\Big( \tau_2\frac{\p}{\p \bar\tau} + \frac{in}{2}\Big), \ee
 whose actions on $\Psi{(m,n)}$ are given by
\be D_m \Psi^{(m,n)} \rightarrow \Psi^{(m+1,n-1)}, \quad \bar{D}_n \Psi^{(m,n)} \rightarrow \Psi^{(m-1,n+1)}.\ee

\section{The tree level five and eight graviton amplitudes from multi--gluon amplitudes}

While the structure of the tree level 8 graviton amplitude is directly relevant for our analysis, the structure of the tree level 5 graviton amplitude is needed in section 4 to analyse the source terms.    
These multi--graviton amplitudes can be calculated directly using the KLT relations~\cite{Kawai:1985xq}. We denote the 5 and 8 graviton amplitudes as
\bea \label{short1}
A^{(5)} (\alpha') &\equiv&  A^{(5)}_{\mu_1\nu_1 \ldots \mu_5\nu_5} (\alpha';k_i)\zeta^{\mu_1\nu_1}_1 \ldots \zeta^{\mu_5\nu_5}_5, \non \\ 
A^{(8)} (\alpha') &\equiv&  A^{(8)}_{\mu_1\nu_1 \ldots \mu_8\nu_8} (\alpha';k_i)\zeta^{\mu_1\nu_1}_1 \ldots \zeta^{\mu_8\nu_8}_8,\eea
respectively, where $\zeta^{\mu\nu}_i$ ($i=1,\ldots,5$ and $i=1,\ldots,8$ for the two cases) is the polarization tensor of the $i$--th graviton carrying momentum $k_i$. Since any closed string tree level amplitude is given as sums of squares of color ordered open string amplitudes along with certain momenta dependent sine factors, we need to know the expressions for the color ordered 5 and 8 gluon tree level amplitudes in the type I theory. We denote them as
\bea \label{short2}
A^{(5)}_{op} (\alpha';a,b,c,d,e) &\equiv& A^{(5)}_{op\mu_1 \ldots \mu_5} (\alpha';k_i;a,b,c,d,e) e^{\mu_1}_1 \ldots e^{\mu_5}_5 , \non \\
A^{(8)}_{op} (\alpha';a,b,c,d,e,f,g,h) &\equiv& A^{(8)}_{op\mu_1 \ldots \mu_8} (\alpha';k_i;a,b,c,d,e,f,g,h) e^{\mu_1}_1 \ldots e^{\mu_8}_8,\eea
where $e^{\mu}_i$ ($i=1,\ldots,5$ and $i=1,\ldots,8$ for the two cases) is the polarization vector of the $i$--th gluon carrying momentum $k_i$. The color ordering is specified by the ordered sequences $a,\ldots,e$ and $a, \ldots, h$. In the expressions below, we have that
\be \zeta^{\mu\nu}_i = e^{\mu}_i \bar{e}^{\nu}_i,\ee
where $e^{\mu}_i$ and $\bar{e}^{\nu}_i$ are the gluon polarization vectors for the world sheet left and right movers respectively. 

For brevity, we introduce the notation
\bea {\rm sin} \Big(\frac{\alpha'\pi}{2} k_i \cdot k_j\Big) \equiv [i,j],  \quad {\rm sin}\Big(\frac{\alpha'\pi}{2} k_i \cdot (k_j + k_l)\Big) \equiv [i,j+l], \non \\  {\rm sin}\Big(\frac{\alpha'\pi}{2} k_i \cdot (k_j + k_l + k_r)\Big) \equiv [i,j+l+r].\eea

For the 5 graviton amplitude, the relation between the open and closed string amplitudes is given by~\cite{Kawai:1985xq}
\bea \label{5long} \pi^2 A^{(5)} (\alpha') =  \Big[ [1,2][3,4] A^{(5)}_{op} (\alpha'/4;1,2,3,4,5) \bar{A}^{(5)}_{op} (\alpha'/4;2,1,4,3,5) \non \\ + [1,3][2,4] A^{(5)}_{op} (\alpha'/4;1,3,2,4,5) \bar{A}^{(5)}_{op} (\alpha'/4;3,1,4,2,5)\Big] .\eea

For the 8 graviton amplitude, the relation between the open and closed string amplitudes involves $12 \cdot 5!$ terms~\cite{Kawai:1985xq}. We shall write down 12 terms explicity, while the rest are related by permutation symmetry as mentioned below.  
Thus the 8 graviton amplitude is given by
\bea \label{long}
\pi^5 A^{(8)} (\alpha') &=&  -[1,2] [6,7] A^{(8)}_{op} (\alpha'/4;1,2,3,4,5,6,7,8)  \times \non \\&&\Big[
[1,3] [4,7] [5,7] \bar{A}^{(8)}_{op} (\alpha'/4;2,3,1,7,4,5,6,8) \non \\ 
&&+[1,3] [4,7] [5,6+7] \bar{A}^{(8)}_{op} (\alpha'/4;2,3,1,7,4,6,5,8)  \non \\
&&+[1,3][5,7][4,5+7]\bar{A}^{(8)}_{op} (\alpha'/4;2,3,1,7,5,4,6,8)  \non \\
&&+[1,3][5,7][4,5+6+7]\bar{A}^{(8)}_{op} (\alpha'/4;2,3,1,7,5,6,4,8)  \non \\
&&+[1,3][4,6+7][5,6+7]\bar{A}^{(8)}_{op} (\alpha'/4;2,3,1,7,6,4,5,8)  \non \\
&&+[1,3][5,6+7][4,5+6+7]\bar{A}^{(8)}_{op} (\alpha'/4;2,3,1,7,6,5,4,8)  \non \eea
\bea
&&+[4,7][5,7][3,2+1]\bar{A}^{(8)}_{op} (\alpha'/4;3,2,1,7,4,5,6,8)  \non \\
&&+[4,7][3,2+1][5,6+7]\bar{A}^{(8)}_{op} (\alpha'/4;3,2,1,7,4,6,5,8)  \non \\
&&+[5,7][3,2+1][4,5+7]\bar{A}^{(8)}_{op} (\alpha'/4;3,2,1,7,5,4,6,8)  \non \\
&&+[5,7][3,2+1][4,5+6+7]\bar{A}^{(8)}_{op} (\alpha'/4;3,2,1,7,5,6,4,8)  \non \\
&&+[3,2+1][4,6+7][5,6+7]\bar{A}^{(8)}_{op} (\alpha'/4;3,2,1,7,6,4,5,8) \non \\ 
&&+[3,2+1][5,6+7][4,5+6+7]\bar{A}^{(8)}_{op} (\alpha'/4;3,2,1,7,6,5,4,8)  \Big] \non \\
&&+{\rm other~ permutations ~of} ~ 2 3 4 5 6. \eea
The permutations of $23456$ in $A^{(8)}_{op}$ yield the $5!$ terms, and in each such sequence there are 12 terms. In \C{long}, we have written the 12 terms corresponding to the specific ordering $23456$ as is evident from $A^{(8)}_{op} (\alpha'/4;1,2,3,4,5,6,7,8)$ in the very first line.

\section{The structure of the $\alpha'$ expansion of the tree level five and eight graviton amplitudes}

In order to calculate multi--graviton amplitudes, it is quite convenient to use the KLT relations~\cite{Kawai:1985xq}. Thus one needs to calculate multi--gluon amplitudes first. These multi--gluon amplitude calculations not only have significance as worldsheet scattering amplitudes, but are also important from the point of view of obtaining the effective action for Yang--Mills fields, and have been widely investigated. 

From the point of view of calculating the effective action, this has led to results that generalize the abelian Born--Infeld action. Various techniques have been developed to calculate these interactions in the effective action at low orders in $\alpha'$, for example: using deformations of BPS solutions~\cite{Koerber:2001uu,Koerber:2001hk,Koerber:2002zb}, directly using supersymmetry~\cite{Collinucci:2002ac,Drummond:2003ex}, pure spinor cohomology techniques~\cite{Cederwall:2001dx,Howe:2010nu}, and vanishing of the $(e \cdot k)^N$ term in the $N$--gluon tree amplitude~\cite{Barreiro:2012aw}, where $e_\mu$ is the polarization vector. Of course, yet another method is to obtain the effective action using scattering amplitudes, which is the one we shall use for constructing the type II action. 

These disc amplitudes have been considered in detail with an aim of calculating the effective action. The structure of the detailed nature of these amplitudes for more than 4 external states are along the lines of our calculations. One of the primary motivations for calculating these amplitudes has been to obtain the structure of field theory amplitudes in the $\alpha' \rightarrow 0$ limit directly from string theory, and also to obtain the $\alpha'$ expansion of these amplitudes in general. These issues have been discussed in~\cite{Medina:2002nk,Barreiro:2005hv,Oprisa:2005wu,Stieberger:2006te,Stieberger:2007jv,Stieberger:2007am,Mafra:2009bz,Stieberger:2009rr,Mafra:2010gj,Mafra:2011nv,Mafra:2011nw}.  

In this section, we perform the $\alpha'$ expansion of the 5 and 8 graviton amplitudes at tree level which lead to the $\mathcal{R}^5$ and $\mathcal{R}^8$ interactions respectively in the effective action, at the linearized level. We shall see that the structure is consistent with the assumption that these interactions are in the same supermultiplet as the $D^2 \mathcal{R}^4$ (which vanishes on--shell) and $D^8 \mathcal{R}^4$ interactions respectively.  
We also briefly discuss the structure of the 6 and 7 graviton amplitudes which lead to the $\mathcal{R}^6$ and $\mathcal{R}^7$ interactions respectively at the linearized level, which is consistent with the assumption that they lie in the same supermultiplet as the the $D^4 \mathcal{R}^4$ and $D^6 \mathcal{R}^4$ interactions respectively. 

To calculate the 5 and 8 graviton amplitudes, we shall use \C{5long} and \C{long} respectively, and hence we need the expressions for the the color ordered 5 and 8 gluon amplitudes in type I string theory. All the calculations are done in the RNS formalism. 

The gluon vertex operators are inserted on the boundary of the disk, which is conformally mapped to the upper half plane. The gluon vertex operator with polarization $e_\mu$ and momentum $k_\mu$ in the 0 picture is given by
\be V^{(0)} = e_\mu (i \p X^\mu + 2 \alpha' k \cdot \psi \psi^\mu ) e^{i k\cdot X},\ee  
and in the $-1$ picture is given by
\be V^{(-1)} = e_\mu e^{-\varphi} \psi^\mu e^{ik\cdot X},\ee
where $\varphi$ is the chiral scalar coming from bosonizing the $\beta-\gamma$ CFT~\cite{Friedan:1985ge}. The momentum and polarization vector satisfy the masslessness and transversality conditions
\be k^2 = 0, \quad e \cdot k =0\ee
respectively. 

For the color ordered N gluon amplitude where the vertex operator $V_N$ corresponds to inserting the N--th gluon on the boundary of the worldsheet at $y_N$, using the $SL(2,\mathbb{R})$ symmetry, we fix the vertex operators $V_1$, $V_{N-1}$ and $V_N$ to be at $0$, $1$ and $\infty$ respectively, and integrate over the rest. The vertex operators $V_{1}$ and $V_{N-1}$ are taken to be in the $-1$ picture while the rest are in the 0 picture~\cite{Friedan:1985ge}. Also we shall not keep track of various overall numerical factors. 

\subsection{The five gluon disc amplitude}

We should note that the 5 gluon as well as the 5 graviton amplitudes have been understood, and the discussion below is along the lines of the existing literature. We have been somewhat detailed in our discussion, because we want to highlight the pole structure in the $\alpha'$ expansion, and see how the coefficients of fixed transcendentality arise. More importantly, this discussion will be generalized to the higher point cases which are considerably more complicated, and thus it is very useful to see the analogous structures for this amplitude.

We consider the color ordered 5 gluon disc amplitude for
\be y_1 < y_2 < y_3 < y_4 < y_5 ,\ee
given by
\be \label{5p} A^{(5)}_{op} (\alpha';12345) = e^{-\phi} \int_0^1 d y_3 \int_0^{y_3} d y_2\langle V^{(-1)}_1 (0) V^{(0)}_2 (y_2) V^{(0)}_3 (y_3) V^{(-1)}_4 (1) V^{(0)}_5 (\infty) \rangle. \ee

It should be noted that various interactions in the low energy multi--gluon effective action can be obtained from \C{5p} on making an $\alpha'$ expansion. These include interactions of the form $F^4$, $D^2 F^4, \ldots$ which involve 5 gluons at the non--linear level, and 
the quintic $F^5$ interaction which has been analyzed in detail~\cite{Kitazawa:1987xj,Medina:2002nk,Barreiro:2005hv,Mafra:2009bz,Boels:2013jua} by calculating the 5 point amplitude, and also using supersymmetric techniques in~\cite{Koerber:2001uu,Koerber:2001hk,Collinucci:2002ac}.

 Our aim is not to investigate the full spacetime structure of the amplitude, but to find the coefficient of the $\mathcal{R}^5$ term starting from \C{5p}. Thus to keep things as simple as possible, we only look at those terms in \C{5p} that are of the form $(e_2 \cdot e_3)(e_4 \cdot e_5)$. 
Thus keeping only the $(e_2 \cdot e_3)(e_4 \cdot e_5)$ terms in \C{5p}, we get that
\bea \label{5op}
A^{(5)}_{op} (\alpha';12345) = (2\alpha')^2 e^{-\phi} (e_2 \cdot e_3)(e_4 \cdot e_5)(I_1 + I_2 + I_3) + \ldots,\eea
where
\bea \label{intthree}
I_1 &=& -\alpha' s_{35} (k_2 \cdot e_1) \int_0^1 dy_3 \int_0^{y_3} dy_2 y_2^{-\alpha' s_{12} -1} y_3^{-\alpha' s_{13}} y_{32}^{-\alpha' s_{23} -1} (1-y_2)^{-\alpha' s_{24}} (1-y_3)^{-\alpha' s_{34}}, \non \\ I_2 &=& \alpha' s_{25} (k_3 \cdot e_1) \int_0^1 dy_3 \int_0^{y_3} dy_2 y_2^{-\alpha' s_{12} } y_3^{-\alpha' s_{13}-1} y_{32}^{-\alpha' s_{23} -1} (1-y_2)^{-\alpha' s_{24}} (1-y_3)^{-\alpha' s_{34}}, \non \\ I_3 &=& -(1+\alpha' s_{23}) (k_5 \cdot e_1) \int_0^1 dy_3 \int_0^{y_3} dy_2 y_2^{-\alpha' s_{12} } y_3^{-\alpha' s_{13}} y_{32}^{-\alpha' s_{23} -2} (1-y_2)^{-\alpha' s_{24}} (1-y_3)^{-\alpha' s_{34}},\non \\ \eea
and the generalized Mandelstam variables are defined by
\be s_{ij} = -(k_i + k_j)^2 = - 2 k_i \cdot k_j, \quad (i \neq j).\ee
We expand the integrals in \C{intthree} in powers of $\alpha'$. We outline some of the steps in the calculation which helps to see the structure of the singularities from the various poles in the amplitude, which is crucial for our purposes. The analysis follows~\cite{Medina:2002nk,Barreiro:2005hv}. 

The expression for $I_1$ involves the integral
\bea \label{long1}
 J_1 &= &\int_0^1 dy_3 \int_0^{y_3} dy_2 y_2^{-\alpha' s_{12} -1} y_3^{-\alpha' s_{13}} y_{32}^{-\alpha' s_{23} -1} (1-y_2)^{-\alpha' s_{24}} (1-y_3)^{-\alpha' s_{34}} \non \\ &=& B(-\alpha' a, 1-\alpha' b) \Big( B(-\alpha' c, 1-\alpha' d) + B(1-\alpha' c, -\alpha'd) \Big)
%\Gamma(1-\alpha' c) \Gamma(1-\alpha' d)}{\Gamma\Big(1-\alpha'(a+b)\Big) \Gamma\Big(1-\alpha%' (c+ d)\Big)}
\non \\ &&+\int_0^1 dx x^{-\alpha' a-1} (1-x)^{-\alpha' b} \int_0^1 du u^{-\alpha' c-1} (1-u)^{-\alpha' d-1} \Big( (1-ux)^{-\alpha'f }-1\Big),\non
%\\ &&-\int_0^1 dx x^{-\alpha' a} (1-x)^{-\alpha' b} \int_0^1 du u^{-\alpha' c} (1-u)^{-\alp%ha' d-1} (1-ux)^{-\alpha'f -1},
\\ \eea
where
\be a = s_{12} + s_{13} + s_{23}, \quad b= s_{34}, \quad c= s_{12}, \quad d= s_{23}, \quad f = s_{24}.\ee 
We have also used the definiton for the Euler beta function
\be \label{EB} B(x,y) = \frac{\Gamma (x) \Gamma (y)}{\Gamma (x+y)}.\ee  
While the term involving \C{EB} in \C{long1} involves the simple poles, the others do not.
On using the integrals in appendix E and
\be \label{relgamma}
{\rm ln} ~\Gamma (1-z) = \gamma z+ \sum_{n=2}^\infty \frac{\zeta (n)}{n} z^n,\ee 
where $\gamma$ is the Euler--Mascheroni constant, the integral in \C{long1} is equal to
\bea \label{fin1}
J_1 &=& \frac{1}{\alpha'^2 a} \Big( \frac{1}{c} + \frac{1}{d}\Big) -\frac{\zeta (2)}{a} \Big( \frac{1}{c} + \frac{1}{d} \Big)(ab+ cd)-\frac{\alpha' \zeta(3)}{a} \Big( \frac{1}{c} + \frac{1}{d} \Big) \Big( ab(a+b) + cd(c+d)\Big) \non \\ && - \alpha' \zeta (3)f + \alpha'^2 \zeta(4) \Big[ \frac{f}{4} \Big(-a+31 b+ 11c + 17d + 13f\Big) \non \\ &&- \frac{1}{a} \Big(\frac{1}{c} + \frac{1}{d} \Big) \Big( ab(a^2 + b^2) + cd(c^2 + d^2) +\frac{a^2 b^2 + c^2 d^2}{4}-\frac{5}{2} abcd \Big)\Big]+  O(\alpha'^3).\non \\ \eea
The integral needed to evaluate $I_2$ is
\bea \label{long2}
 J_2 &=& \int_0^1 dy_3 \int_0^{y_3} dy_2 y_2^{-\alpha' s_{12} } y_3^{-\alpha' s_{13}-1} y_{32}^{-\alpha' s_{23} -1} (1-y_2)^{-\alpha' s_{24}} (1-y_3)^{-\alpha' s_{34}} \non \\ &=& B(-\alpha' a, 1-\alpha' b) B(1-\alpha' c, -\alpha'd)
%\frac{\Gamma(-\alpha'a)\Gamma(1-\alpha' b) \Gamma(1-\alpha' c) \Gamma(-\alpha' d)}{\Gamma\B%ig(1-\alpha'(a +b)\Big) \Gamma\Big(1-\alpha' (c+ d)\Big)}
\non \\ &&+\int_0^1 dx x^{-\alpha' a-1} (1-x)^{-\alpha' b} \int_0^1 du u^{-\alpha' c} (1-u)^{-\alpha' d-1} \Big( (1-ux)^{-\alpha'f }-1\Big).\non \\
%\non \\ &&-\int_0^1 dx x^{-\alpha' a} (1-x)^{-\alpha' b} \int_0^1 du u^{-\alpha' c +1} (1-u%)^{-\alpha' d-1} (1-ux)^{-\alpha'f -1}.
\eea
Again, the term involving \C{EB} contains the poles, and thus we get that
\bea J_2 &=& \frac{1}{\alpha'^2 ad} - \zeta (2) \frac{(ab+cd)}{ad} - \frac{\zeta (3)}{ad} \alpha' \Big( ab(a+b) + cd(c+d)\Big) - 2\alpha' \zeta (3) f \non \\ &&+ \alpha'^2\zeta(4)  f \Big( -\frac{5}{4}a + \frac{13}{2} b + \frac{7}{4} c + 3(d +f) \Big) \non \\ &&- \alpha'^2\frac{\zeta(4)}{ad} \Big(  ab(a^2 + b^2) + cd(c^2 + d^2) +\frac{a^2 b^2 + c^2 d^2}{4}-\frac{5}{2} abcd\Big) + O(\alpha'^3). \eea

Finally, the integral involved in $I_3$ is
\bea J_3 &=& (1+\alpha' s_{23}) \int_0^1 dy_3 \int_0^{y_3} dy_2 y_2^{-\alpha' s_{12} } y_3^{-\alpha' s_{13}} y_{32}^{-\alpha' s_{23} -2} (1-y_2)^{-\alpha' s_{24}} (1-y_3)^{-\alpha' s_{34}} \non \\ &=& -\alpha' s_{13} J_2 + \alpha' s_{34} \hat{J}_1, \eea
where $\hat{J}_1$ is $J_1$ with
\be s_{12} \leftrightarrow s_{34}, \quad s_{13} \leftrightarrow s_{24}, \quad s_{23} \leftrightarrow s_{23},\ee
and so has its pole structure completely determined by $J_1$.
Thus, by the steps mentioned before, one can determine $J_3$ easily.

The exact details of $J_1, J_2$ and $\hat{J}_1$ are not relevant to us. However, note that they all have the schematic structure
\be  \frac{1}{\alpha'^2 s^2} + \zeta (2) + \zeta (3) \alpha' s + \zeta (4) \alpha'^2 s^2 + O(\alpha'^3),\ee
where $s_{ij} \sim s$ is a typical Mandelstam variable. Thus, for any color ordering of the external gluons, \C{5op} must have the general structure
\be \label{colord}
A^{(5)}_{op} (\alpha';\ldots) \sim e^{-\phi} (e\cdot e)^2 (k\cdot e) \Big( \frac{1}{\alpha' s} + \zeta (2) \alpha' s + \zeta (3) (\alpha' s)^2 + \zeta (4) (\alpha' s)^3 + O(\alpha'^4)\Big)+\ldots,\ee
where we have dropped overall factors of $\alpha'$, and the various momenta depend on the color ordering. 
 
The various terms in \C{colord} can be used the deduce various terms in the effective action. The pole term in \C{colord} comes from the ${\rm tr} F^2$ Yang--Mills term. The $\zeta (2)$ term comes from Feynman diagrams with a vertex involving $\zeta(2) ({\rm tr} F^4 + {\rm tr} (F^2)^2)$, involving both pole and contact terms\footnote{The 4 gluon color ordered amplitude in the $s-u$ channel is given by
\bea \label{4pt}
A^{(4)}_{op} (\alpha; 1234) &=& e^{-\phi}\frac{\Gamma(-\alpha' s)\Gamma(-\alpha' u)}{\Gamma(1+ \alpha' t)}t_8 {\rm tr} F^4 \non \\ &=& e^{-\phi}\Big(\frac{1}{\alpha'^2 su} -\zeta (2) + \zeta (3) \alpha' t -\frac{\zeta (4)}{4} \alpha'^2(4(s^2 + u^2) + su)+ O(\alpha'^3)\Big) t_8 {\rm tr} F^4.\eea}. The $\zeta (3)$ term has a similar origin from vertices of the form $\zeta (3){\rm tr} D^2  F^4$ (see \C{4pt}) and $\zeta (3) {\rm tr}F^5$. Finally the $\zeta (4)$ term is a contact term coming from a $\zeta (4) {\rm tr}D^2 F^5$ term in the effective action~\cite{Barreiro:2005hv}. One obtains an infinite number of terms in the effective action this way. 

\subsection{The coefficient of the $\mathcal{R}^5$ term}
 
Now we can determine the structure of the $\mathcal{R}^5$ interaction from \C{5long}. From the discussion above, using\footnote{The amplitude actually yields $F^5$ at the linearized level.}
\be A^{(5)}_{op} \sim e^{-\phi}F^5 \Big( \frac{1}{(\alpha' s)^3} +\frac{\zeta(2)}{\alpha' s} +\zeta (3) +\zeta (4) \alpha' s + O(\alpha'^2)\Big),  \ee
and
\be -{\rm sin} (-\pi \alpha' s) = \pi \alpha' s \Big( 1 + \zeta (2) (\alpha' s)^2 + \frac{3}{4} \zeta (4) (\alpha' s)^4 + O(\alpha'^5)\Big),\ee
we see that 
\be \label{5exp}
A^{(5)}_{cl} \sim e^{-2\phi} \mathcal{R}^5 \Big( \frac{1}{(\alpha' s)^4} + \frac{\zeta(3)}{\alpha' s} + O(\alpha')\Big),\ee
where $\mathcal{R}^5 \sim (F^5)^2$. In \C{5exp}, the leading pole term comes from Einstein gravity, while the $\zeta (3)$ term comes from Feynman diagrams involving a $\zeta (3) \mathcal{R}^4$ vertex at the linearized level\footnote{The 4 graviton amplitude is given by
\bea \label{grav4}
A^{(4)}_{cl} &=& -e^{-2\phi}\frac{\Gamma(-\alpha's/4)\Gamma(-\alpha' t/4)\Gamma(-\alpha' u/4)}{\Gamma(1+\alpha's/4)\Gamma(1+\alpha't/4)\Gamma(1+\alpha' u/4)}t_8 t_8 R^4 \non \\ &=& e^{-2\phi}\Big( \frac{64}{\alpha'^3 stu} + 2 \zeta (3) + \frac{\zeta (5)}{16} \alpha'^2 (s^2 + t^2 + u^2) + \frac{\zeta (3)^2}{32} \alpha'^3 stu+ \frac{\zeta(7)}{512} (s^2 + t^2 + u^2)^2\alpha'^4 \non \\ &&+ O(\alpha'^5)\Big) t_8 t_8 R^4.\eea}. We have dropped the terms proportional to $(\alpha s)^{-2}$ and $O(1)$ in \C{5exp} which involved various contributions involving $\zeta (2)$ and $\zeta (4)$ respectively. The pole term involving $\zeta (2)$ must vanish because there is no interaction at lower orders in the derivative expansion which could have given rise to such a term. The contact term involving $\zeta (4)$ has been set to zero because the $\mathcal{R}^5$ term should be in the same supermultiplet as $D^2 \mathcal{R}^4$ which vanishes on--shell. This vanishing has indeed been observed in~\cite{Stieberger:2009rr}\footnote{This is alo consistent with the analysis of the 5 loop beta function in~\cite{Grisaru:1986wj}.}.

By the same reasoning, we expect the $\mathcal{R}^5$ term in the effective action to vanish to all orders in perturbation theory, as well as non--perturbatively. In fact, the one loop vanishing of the $\mathcal{R}^5$ interaction has been noted in~\cite{Richards:2008jg}. It would be interesting to directly prove the vanishing of the $\mathcal{R}^5$ interaction. 

\subsection{Brief schematics of the $\mathcal{R}^6$ and $\mathcal{R}^7$ coefficients}

Though we have not done any detailed analysis of terms in the effective action at higher orders in the $\alpha'$ expansion that are obtained from the 5 gluon amplitude, we expect transcendentality of the Riemann zeta functions to yield the general structure exactly along the lines of the 5 gluon disc amplitude analysis done above. Thus, we expect terms like 
\be \label{morealpha}
\Big(\zeta (5) + \zeta (2) \zeta (3)\Big){\rm tr} D^4 F^5, \quad \Big(\zeta(6) + \zeta (3)^2\Big){\rm tr}D^6 F^5  , \quad \Big(\zeta(7) + \zeta(2) \zeta (5) + \zeta (3) \zeta (4)\Big){\rm tr} D^8 F^5\ee
and so on in the effective action, where each individual term at a fixed order in the $\alpha'$ expansion has a different spacetime structure. In support of this statement, we have evaluated only a few integrals that arise in calculating these amplitudes, which are given in \C{d4f5}, \C{d6f5} and \C{d8f5}. They yield the terms of the type mentioned above, and no others.

Because of maximal supersymmetry, as for the multi--graviton amplitudes, we expect the ${\rm tr} F^6, {\rm tr}F^7, {\rm tr} F^8$ and ${\rm tr} F^9$ terms to be in the same supermultiplet as ${\rm tr} D^2 F^5, {\rm tr} D^4 F^5, {\rm tr} D^6 F^5$ and ${\rm tr} D^8 F^5$ terms respectively, and so on. This is alo expected for the non--abelian theory because one can use $[D,D] F = F^2$ repeatedly, so that the definition of the coefficient of any of these operators simply by itself is ambiguous. Thus, for example, the ${\rm tr} F^6, {\rm tr} D^2 F^5$ and ${\rm tr} D^4 F^4$ couplings should be the same, not only at tree level but even beyond.
Thus \C{morealpha} provides the coefficients of several other terms in the effective action that are related to these interactions by supersymmetry. 

What are the implications of this for the $\mathcal{R}^6$ and $\mathcal{R}^7$ interactions? Based on the above arguments, the 6 gluon amplitude must take the form
\be A^{(6)}_{op} \sim e^{-\phi} F^6 \Big( \frac{1}{(\alpha' s)^4} +\frac{\zeta(2)}{(\alpha's)^2 }+ \frac{\zeta (3)}{\alpha' s} +\zeta (4) + \Big(\zeta (5) +\zeta (2)\zeta (3)\Big) \alpha's + O(\alpha'^2)\Big), \ee
where the most singular term is the Yang--Mills contribution, the $\zeta (2)$ term involves the ${\rm tr}F^4$ vertex, the $\zeta (3)$ term involves either the ${\rm tr} F^5$ or the ${\rm tr} D^2 F^4$ vertex, the $\zeta (4)$ term involves the new ${\rm tr} F^6$ term, as well as the square of the ${\rm tr} F^4$ vertex, and the ${\rm tr}D^2 F^5$ contact term, while the $\zeta (5) +\zeta (2) \zeta (3)$ term involves the new ${\rm tr} D^2 F^6$ vertex, and the product of the $\zeta (2) {\rm tr} F^4$ and $\zeta(3){\rm tr} D^2 F^4$ vertex, and so on. By new vertices in the effective action, we mean new interactions that arise at this order in the $\alpha'$ expansion.

Then, using the KLT relation for the 6 graviton amplitude given by
\be A^{(6)}_{cl} \sim  ({\rm sin} \pi \alpha' s)^3 (A^{(6)}_{op})^2,\ee
the 6 graviton amplitude must take the form
\be \label{6grav}
A^{(6)}_{cl} \sim e^{-2\phi} \Big( \frac{1}{(\alpha's)^5} + \frac{\zeta(3)}{(\alpha's)^2} +\zeta (5) + O(\alpha')\Big) \mathcal{R}^6.\ee
As in the analysis before, the three terms are the contributions from Einstein gravity, the $\zeta(3) \mathcal{R}^4$ term, the $\zeta(5) D^4 \mathcal{R}^4$ term (see \C{grav4}) and a new $\zeta (5) \mathcal{R}^6$ interaction. We have dropped terms of the form $\zeta(2)/(\alpha' s)^3$ and $\zeta (2) \zeta (3)$ in \C{6grav}, because the first one does not arise from any vertex, and the second because we have assumed that the $\mathcal{R}^6$ and $D^4\mathcal{R}^4$ interactions are in the same supermultiplet, and have the same couplings. 
The vanishing of the $\zeta(2) \zeta (3)$ term has been observed in~\cite{Stieberger:2009rr}.

Thus this yields an interaction
\be \zeta (5) \int d^{10} x \sqrt{-g} e^{-2\phi} \mathcal{R}^6\ee
in the low energy effective action.

Running through the same logic, the 7 gluon amplitude must take the form
\be A^{(7)}_{op} \sim e^{-\phi} F^7 \Big( \frac{1}{(\alpha' s)^5} + \frac{\zeta (2)}{(\alpha' s)^3} + \frac{\zeta (3)}{(\alpha' s)^2} + \frac{\zeta (4)}{\alpha' s} + \Big(\zeta (5) + \zeta (2) \zeta (3)\Big)+\Big(\zeta (6) + \zeta (3)^2\Big) \alpha' s + O(\alpha'^2)\Big),\ee
leading to
\be A^{(7)}_{cl} \sim e^{-2\phi} \mathcal{R}^7 \Big( \frac{1}{(\alpha' s)^6}+\frac{\zeta(3)}{(\alpha' s)^3} +\frac{\zeta(5)}{\alpha' s}+ \zeta(3)^2 + O(\alpha')\Big)\ee
on using the KLT relation
\be A^{(7)}_{cl} \sim ({\rm sin} \pi \alpha' s)^4 (A^{(7)}_{op})^2,\ee
and dropping zeta functions of even transcendentality, based on supersymmetry\footnote{This has been conjectured to be true upto a certain order in the momentum expansion in~\cite{Schlotterer:2012ny}, beyond what we are interested in.}. 
This leads to an interaction
\be \zeta (3)^2 \int d^{10} x \sqrt{-g} e^{-2\phi} \mathcal{R}^7\ee
in the low energy effective action.

\subsection{The eight gluon disc amplitude}

Let us now consider the 8 gluon disc amplitude. We consider the color ordered amplitude for
\be y_1 < y_2 < y_3 < y_4 < y_5 < y_6 < y_7 < y_8 ,\ee 
given by 
\bea \label{pt8}
&& A^{(8)}_{op} (\alpha';12345678) \non \\ &&=  e^{-\phi}\int_0^1 dy_6 \int^{y_6}_0 dy_5\int^{y_5}_0 dy_4 \int^{y_4}_0 dy_3 \int_0^{y_3} dy_2  \langle V_1^{(-1)} (0) \prod_{i=2}^6 V_i^{(0)} (y_i)V_7^{(-1)} (1) V_8^{(0)} (\infty) \rangle.\non \\ \eea
Among the very large number of possible contractions that arise, we only look at the $(e_1 \cdot e_2)(e_3 \cdot e_4)(e_5 \cdot e_6)(e_7 \cdot e_8)$ term\footnote{Thus the only contributions from \C{pt8} are terms with 4, 2 and 0 factors of $\p X$ in the correlator.}. Thus, defining
\be \int [dy] \equiv \int_0^1 dy_6 \int^{y_6}_0 dy_5\int^{y_5}_0 dy_4 \int^{y_4}_0 dy_3 \int_0^{y_3} dy_2 ,\ee
and 
\be \Lambda \equiv \Big[ \prod_{i=2}^6 y_i^{-\alpha' s_{1i}} (1-y_i)^{-\alpha' s_{i7}} \Big] \Big[ \prod_{i,j=2,\ldots,6;i>j} y_{ij}^{-\alpha' s_{ij}}\Big],\ee
we get that
\be A^{(8)}_{op} (\alpha';12345678) =  (2\alpha')^3 e^{-\phi} (e_1 \cdot e_2)(e_3 \cdot e_4)(e_5 \cdot e_6)(e_7 \cdot e_8)\sum_{i=1}^{15} K_i +\ldots,\ee
where the 15 $K_i$ integrals are given by
\bea \label{int10}
K_1 = \alpha'^3 s_{23} s_{45} s_{68} \int  \frac{[dy]}{y_2 y_{32} y_{43} y_{54} y_{65}}\Lambda, \quad K_2 = -\alpha'^3 s_{23} s_{46} s_{58}\int  \frac{[dy]}{y_2 y_{32} y_{43} y_{65} y_{64}}\Lambda,\non \\ K_3 = -\alpha'^3 s_{24} s_{35} s_{68} \int  \frac{[dy]}{y_2 y_{42} y_{43} y_{53} y_{65} }\Lambda, \quad K_4 = \alpha'^3 s_{24} s_{36} s_{58} \int \frac{[dy]}{y_2 y_{42} y_{43} y_{63} y_{65} }\Lambda, \non \\ K_5 = -\alpha'^3 s_{25} s_{36} s_{48} \int \frac{[dy]}{y_2 y_{43} y_{52} y_{63} y_{65} } \Lambda, \quad K_6 = \alpha'^3 s_{25} s_{38} s_{46} \int \frac{[dy]}{y_2 y_{43} y_{52} y_{64} y_{65} }\Lambda, \non \eea
\bea K_7 = \alpha'^3 s_{26} s_{35} s_{48} \int \frac{[dy]}{y_2 y_{43} y_{53} y_{62}y_{65}}\Lambda, \quad K_8 = - \alpha'^3 s_{26} s_{38} s_{45} \int \frac{[dy]}{y_2 y_{43} y_{54} y_{62} y_{65}}\Lambda, \non \\ K_9 = -\alpha'^3 s_{28} s_{35} s_{46}\int \frac{[dy]}{y_2 y_{43} y_{53} y_{64} y_{65}}\Lambda, \quad K_{10} = \alpha'^3 s_{28} s_{36} s_{45} \int \frac{[dy]}{y_2 y_{43} y_{54} y_{63} y_{65} }\Lambda , \non \\ \eea
and
\bea  \label{int15}
K_{11} = \alpha'^2 s_{25} s_{68} (1+\alpha' s_{34}) \int \frac{[dy]}{y_2 y_{43}^2 y_{52} y_{65}}\Lambda, \non \\ K_{12} = -\alpha'^2 s_{26} s_{58} (1+\alpha' s_{34}) \int \frac{[dy]}{y_2 y_{43}^2 y_{62} y_{65}}\Lambda, \non \\ K_{13} =  \alpha'^2 s_{23} s_{48} (1+\alpha' s_{56}) \int \frac{[dy]}{y_2 y_{32} y_{43} y_{65}^2}\Lambda ,\non \\ K_{14} = -\alpha'^2 s_{24} s_{38} (1+\alpha' s_{56}) \int \frac{[dy]}{y_2 y_{43} y_{42} y_{65}^2}\Lambda, \non \\ K_{15} = \alpha' s_{28} (1+\alpha' s_{34}) (1+\alpha' s_{56}) \int \frac{[dy]}{y_2 y_{43}^2 y_{65}^2}\Lambda. \non \\  \eea

The integrals above have a general structure. While the open string propagators in the denominators of $K_1, \cdots, K_8$ in \C{int10} are all of the form (no sums on repeated indices) $y_{ij} y_{jk} y_{kl} y_{lm} y_{mn}$, 
the denominators of $K_9$ and $K_{10}$ are of the form
$y_{ij} y_{kl} y_{km} y_{ln} y_{mn}$.
Also on integrating by parts, each integral in \C{int15} can be expressed as sums of integrals having denominators with 5 distinct propagators. Thus every integral is of the form
\be \label{5dis} (\alpha' s)^3 \int \frac{[dy]}{y_{ij} y_{kl} y_{mn} y_{pq} y_{rs}}\Lambda,\ee
where none of the 5 propagators in the demominator are the same.
For this to work, it is quite crucial that the correct factors of $(1+\alpha' s_{ij})$ have come out in every term in \C{int15}. For example, the integral involved in $K_{11}$ gives    
\be (1+\alpha' s_{34}) \int \frac{[dy]}{y_2 y_{43}^2 y_{52} y_{65}}\Lambda = \int \frac{[dy]}{y_2 y_{43} y_{52} y_{65}} \Big( -\frac{\alpha' s_{14}}{y_4} - \frac{\alpha' s_{24}}{y_{42}} + \frac{\alpha' s_{54}}{y_{54}} +\frac{\alpha' s_{64}}{y_{64}}  + \frac{\alpha' s_{74}}{1-y_4}\Big) \Lambda.\ee

This pattern is true in general. Thus it is enough for our purposes to look at integrals that are of the form \C{5dis}\footnote{The analogs of $K_1, \ldots, K_{10}$ for the 5 point function are $I_1$ and $I_2$, while the analog of $K_{11},\ldots, K_{15}$ is $I_3$.}. While every integral can be analyzed based on the discussion below, we shall consider one such integral in some detail to see the structure, and also analyze the transcendentality of the coefficients that arise. The other integrals all involve integrations of the same type with different choices of $y_{ij}$, and must give the same pole structure and transcendentality. Even though the intermediate steps are quite involved, the final answer has a simple structure dictated by supersymmetry.   

We consider the integral $K_1$ in \C{int10} with the overall momentum factors stripped off. We get that
\bea \label{intl}
L_1 &=& \int \frac{[dy]}{y_2 y_{32} y_{43} y_{54} y_{65}}\Lambda \non \\ 
&=& \int_0^1 dp p^{-\alpha' A-1} (1-p)^{-\alpha' B} \times \non \\ &&\int_0^1 dy y^{-\alpha' C-1}(1-y)^{-\alpha' D-1} (1-yp)^{-\alpha' E} \times \non \\ &&\int_0^1 dw w^{-\alpha' F-1} (1-w)^{-\alpha' G-1} (1-wy)^{-\alpha' H} (1-wyp)^{-\alpha' I} \times \non \\ &&\int_0^1 dx x^{-\alpha' J-1} (1-x)^{-\alpha' K-1} (1-xw)^{-\alpha' L} (1-xwy)^{-\alpha' M} (1-xwyp)^{-\alpha' N} \times \non \\ &&\int_0^1 dz z^{-\alpha' Q-1} (1-z)^{-\alpha' R-1} (1-zx)^{-\alpha' S} (1-zxw)^{-\alpha' T} (1-zxwy)^{-\alpha' U} (1-zxwyp)^{-\alpha' V},\non \\ \eea
where
\bea &&A=s_{12} + s_{13} + s_{23} +s_{14} +s_{24} + s_{34} + s_{15} + s_{25} + s_{35} + s_{45} + s_{16} + s_{26} + s_{36} + s_{46} + s_{56}, \non \\ &&B= s_{67}, \quad C=s_{12} + s_{13} + s_{23} +s_{14} +s_{24} + s_{34} + s_{15} + s_{25} + s_{35} + s_{45}, \non \\ &&D=s_{56},  \quad E=s_{57}, \quad F = s_{12} + s_{13} + s_{23} + s_{14} + s_{24} + s_{34}, \quad G=s_{45}, \quad H=s_{46}, \non \\ &&I=s_{47}, \quad J=s_{12} + s_{13} + s_{23}, \quad K=s_{34}, \quad L=s_{35}, \quad M=s_{36}, \quad N=s_{37}, \non \\ &&Q=s_{12}, \quad R=s_{23}, \quad S=s_{24}, \quad T=s_{25}, \quad U=s_{26}, \quad V=s_{27}.  \non \\ \eea

Let us now analyse the pole structure, and transcendentality in the $\alpha'$ expansion to the required order, of \C{intl}. Simple dimensional analysis of the 8 gluon amplitude shows that the leading pole structure of \C{intl} is of the form $s^{-5}$, where $s$ is a generic Mandelstam variable. To perform the $\alpha'$ expansion of \C{intl}, we keep all factors of the form $\mu$ and $(1-\mu)$ as it is, where $\mu = p,y,w,x,z$, which gives the leading pole contribution. All other factors are of the form $(1-\lambda_1)^{-\lambda_2}$, where $\lambda_1$ involves at least 2 of the integration variables (there are 10 such terms). We write each of them as 
\be \label{trivial} \Big( (1-\lambda_1)^{-\lambda_2} -1\Big) +1.\ee       
Now we can perform a perturbative expansion in $t\equiv (1-\lambda_1)^{-\lambda_2} -1$, which is the $\alpha'$ expansion. We call it the $t$ expansion, and label the contribution to $L_1$ at $O(t^n)$ by $L_1^{(n)}$, which is an infinite series as a perturbative expansion in $\alpha'$ for every $n$. Note that for a fixed $n$, different terms in $L_1^{(n)}$ have different leading pole singularities, in fact all of them do not even have singularities. This is because these different terms involve integrals that effectively reduce to different multi--gluon amplitudes which have distinct momentum dependence.    

To begin with, at $O(t^0)$, there is only one contribution which is obtained by taking only the $+1$ part of \C{trivial} for all the 10 terms, leading to
\bea \label{s-5}
L_1^{(0)} = B(-\alpha' A,1-\alpha'B) \Sigma(-\alpha' C, -\alpha' D)  \Sigma( -\alpha' F, -\alpha' G)\Sigma(-\alpha' J, \alpha' K) \Sigma(-\alpha' Q, -\alpha' R),\non \\ \eea
where we have defined
\be \Sigma (P,Q) \equiv B(P,1+Q) + B(1+P,Q),\ee
which has a simple pole at $P=0$ and at $Q=0$.
Thus, it immediately follows that \C{s-5} has leading singularity of the form $s^{-5}$. In fact, using \C{relgamma}, it follows that
\bea \label{0exp}
L_1^{(0)} &\sim& \frac{1}{(\alpha's)^5} \Big( 1 + \zeta (2) (\alpha's)^2 + \zeta (3) (\alpha's)^3 + \zeta (4) (\alpha's)^4 + \Big(\zeta (5) + \zeta (2) \zeta (3)\Big) (\alpha's)^5 \non \\ &&+ \Big(\zeta (6) + \zeta (3)^2\Big)(\alpha's)^6 + \Big(\zeta (7) + \zeta (2) \zeta (5) + \zeta (3) \zeta (4)\Big) (\alpha's)^7+  O(\alpha'^8)\Big).\non \\ \eea 

Next consider terms at $O(t)$ in the $t$ expansion, which will have subleading singularities compared to $O(s^{-5})$. At $O(t)$, we have to keep any one of the 10 terms $\Big((1-\lambda_1)^{-\lambda_2} -1\Big) \sim O(\alpha')$, while keeping $+1$ for the remaining 9 from \C{trivial}, hence there are 10 contributions. 
We shall report only a few of the calculations, the analysis for the others follows exactly along the same lines. This shows that the pole structure along with transcendentality is obeyed. This also shows that the leading pole structure is different for different terms. 

From \C{intl}, one such contribution is given by\footnote{This contribution has the most singular pole structure among all terms in $L_!^{(1)}$. }
\bea \label{L1}L_1^{(1)} &=&  \Sigma( -\alpha' F, -\alpha' G)\Sigma(-\alpha' J, -\alpha' K) \Sigma(-\alpha' Q, -\alpha' R)\times \non \\ &&\int_0^1 dp p^{-\alpha' A-1} (1-p)^{-\alpha' B} \int_0^1 dy y^{-\alpha' C-1} (1-y)^{-\alpha' D-1} \Big[(1-yp)^{-\alpha' E} -1\Big] , \non \\\eea
which involves only a 5 point amplitude calculation. This 5 point amplitude integral is exactly the same integral as the one in $J_1$ in \C{long1}, and thus the leading singularity of $L_1^{(1)}$ is $O(s^{-2})$. In the $\alpha'$ expansion of \C{L1}, we have to keep terms upto $O(s^2)$, and there are a huge number of terms, all of which involve integrals that can be treated the same way. We write down only a couple of the relevant integrals that arise at each order of transcendentality in appendix E, all the others can be done similarly, and must have the same structure. Thus for this term, we see that
\bea \label{1exp}
L_1^{(1)} &\sim &\frac{1}{(\alpha's)^2} \Big( \zeta (3)  + \zeta (4) s + \Big(\zeta (5) + \zeta (2) \zeta (3)\Big) (\alpha's)^2 + \Big(\zeta (6) + \zeta (3)^2\Big)(\alpha's)^3 \non \\ &&+ \Big(\zeta (7) + \zeta (2) \zeta (5) + \zeta (3) \zeta (4)\Big) (\alpha's)^4+  O(\alpha'^5)\Big).\eea

Another contribution to $L_1^{(1)}$ is given by
\bea &&L_1^{(1)} = \Sigma(-\alpha' J, -\alpha' K) \Sigma(-\alpha' Q, -\alpha' R) \int_0^1 dp p^{-\alpha' A-1} (1-p)^{-\alpha' B} \times \non \\ &&\int_0^1 dy y^{-\alpha' C-1} (1-y)^{-\alpha' D-1} \int_0^1 dw w^{-\alpha'F-1} (1-w)^{-\alpha'G -1} \Big[ (1-wyp)^{-\alpha' I}-1\Big],\non \\\eea
which involves a 6 point amplitude calculation. Of the many integrals which arise in this calculation and also the ones later on, we only mention a few in appendix E, while the others yield answers of the same transcendentality. Thus we get that  
\be \label{zeta4}L_1^{(1)} \sim \frac{\zeta (4)}{\alpha's} + \ldots, \ee
where the remaining terms have the same expansion as in \C{0exp}. A 7 point amplitude which contributes at this order is given by
\bea \label{7p}
 &&L_1^{(1)} = \Sigma(-\alpha' Q, -\alpha' R) \int_0^1 dp p^{-\alpha' A-1} (1-p)^{-\alpha' B} \int_0^1 dy y^{-\alpha' C-1} (1-y)^{-\alpha' D-1} \times \non \\ &&\int_0^1 dw w^{-\alpha'F-1} (1-w)^{-\alpha'G -1} \int_0^1 dx x^{-\alpha' J-1} (1-x)^{-\alpha'K -1} \Big[ (1-pywx)^{-\alpha' N}-1\Big],\non \\\eea
which has no poles, and the leading contribution is proportional to $\zeta (5)$. Let us also mention an 8 point amplitude which contributes at this order, given by
\bea \label{8p}
&&L_1^{(1)} = \int_0^1 dp p^{-\alpha' A-1} (1-p)^{-\alpha' B} \int_0^1 dy y^{-\alpha' C-1} (1-y)^{-\alpha' D-1} \int_0^1 dw w^{-\alpha'F-1} (1-w)^{-\alpha'G -1} \times \non \\ &&\int_0^1 dx x^{-\alpha' J-1} (1-x)^{-\alpha'K -1} \int_0^1 dz z^{-\alpha' Q-1} (1-z)^{-\alpha' R-1} \Big[ (1-pywxz)^{-\alpha' V}-1\Big],\eea
which has no poles, and the leading contribution is $\zeta (6)(\alpha's)$. The subsequent terms in \C{7p} and \C{8p} match the structure in \C{0exp}.

One can now calculate terms in $L_1^{(n)}$ for higher $n$. There are no poles in any of them, and each contribution has a smooth limit as $s\rightarrow 0$. The results are along the lines of what we have discussed above.

\subsection{The coefficient of the $\mathcal{R}^8$ term}

We now analyze the structure of the $\mathcal{R}^8$ interaction. From \C{0exp}, we get that
\bea A^{(8)}_{op} &\sim& \frac{F^8}{(\alpha's)^6} \Big( 1 + \zeta (2) (\alpha' s)^2 + \zeta (3) (\alpha' s)^3 + \zeta (4) (\alpha' s)^4 + (\zeta (5) + \zeta (2) \zeta (3))(\alpha' s)^5  \non \\ &&+ (\zeta (6) + \zeta(3)^2)(\alpha' s)^6 + (\zeta (7) + \zeta (2) \zeta (5) + \zeta (3) \zeta (4))(\alpha' s)^7 + O(\alpha'^8)\Big),\eea
where it is easy to see that the various terms can be interpreted as giving rise to various interactions in the effective action, as discussed before. The new terms are the $(\zeta (6) +\zeta(3)^2){\rm tr}F^6$, and $(\zeta (7) +\zeta (2) \zeta (5) + \zeta (3) \zeta (4)) {\rm tr}D^2 F^6$ contact interactions. On using \C{long}, thus we get that
\bea \label{8ptexp}
A^{(8)}_{cl} \sim \mathcal{R}^8 \Big( \frac{1}{(\alpha' s)^7} + \frac{\zeta (3)}{(\alpha' s)^4} + \frac{\zeta (5)}{(\alpha' s)^2} + \frac{\zeta(3)^2}{(\alpha' s)} + \zeta (7) + O(\alpha')\Big).\eea
Again we have dropped the terms involving zeta functions of even transcendentality. In \C{8ptexp}, the first term is the contribution from Einstein gravity, the $\zeta (3)$ term involves the $\mathcal{R}^4$ vertex, while the $\zeta (5)$ term involves either the $D^4 \mathcal{R}^4$ or the $D^2 \mathcal{R}^5$ vertex. The $\zeta (3)^2$ term involves the $D^6 \mathcal{R}^4$ vertex, as well as the square of the $\mathcal{R}^4$ vertex. Finally, the $\zeta (7)$ term leads to a contact interaction of the form
\be \label{8ptint}\zeta (7) \int d^{10} x \sqrt{-g} e^{-2\phi} \mathcal{R}^8 \ee  
in the effective action.

The structure of multi--graviton amplitudes we have discussed above should generalize to higher point functions, and also to higher orders in the momentum expansion of the amplitudes we have considered. In particular, zeta functions of even transcendentality should not contribute to closed string tree level amplitudes. It would be of interest to prove this assertion.

\section{List of integrals}
\label{listint}

Various integrals are needed to obtain the $\alpha'$ expansion of the 5 and 8 gluon amplitudes. We list them below. In every case, all the relevant integrals at a fixed order in the $\alpha'$ expansion produce Riemann zeta functions of a fixed transcendentality, upto overall numerical factors.

\subsection{Integrals for the five gluon amplitude}

For the 5 gluon amplitude, we list all the integrals need in our analysis. They are
\be \int_0^1 dx\frac{{\rm ln} x}{1-x} = -\zeta (2),\ee
and
\bea \int_0^1 dx \int_0^1 du \frac{{\rm ln} (1-ux)}{x(1-u)} &=& 2 \zeta(3),\non \\
\int_0^1 dx \int_0^1 du \frac{{\rm ln} (1-ux)}{ux(1-u)} &=& \zeta(3),
\eea
as well as
\bea
\int_0^1 dx \int_0^1 du \frac{1}{xu} {\rm ln} (1-ux) {\rm ln} (1-u) &=& \frac{5}{4} \zeta (4), \non \\
\int_0^1 dx \int_0^1 du \frac{1}{xu} {\rm ln} (1-ux) {\rm ln} u &=& \zeta (4), \non \\ 
\int_0^1 dx \int_0^1 du \frac{1}{xu} {\rm ln} (1-ux) {\rm ln} (1-ux) &=& \frac{1}{2} \zeta (4), \non \\
\int_0^1 dx \int_0^1 du \frac{1}{x(1-u)} {\rm ln} (1-ux) {\rm ln} (1-u) &=& 3\zeta (4) , \non \\
\int_0^1 dx \int_0^1 du \frac{1}{x(1-u)} {\rm ln} (1-ux) {\rm ln} u &=& \frac{7}{4} \zeta (4) , \non \\
\int_0^1 dx \int_0^1 du \frac{1}{x(1-u)} {\rm ln} (1-ux) {\rm ln} (1-x) &=& \frac{13}{2} \zeta (4), \non \\
\int_0^1 dx \int_0^1 du \frac{1}{x(1-u)} {\rm ln} (1-ux) {\rm ln} x &=& -\frac{5}{4} \zeta (4) , \non \\
\int_0^1 dx \int_0^1 du \frac{1}{x(1-u)} {\rm ln} (1-ux) {\rm ln} (1-ux) &=& 6 \zeta (4) .
\eea
These integrals follow from the various tables of integrals on using various identities, and we shall outline the details of only one of them. 
We have that
\be \label{demo1}\int_0^1 dx \int_0^1 du \frac{1}{xu} {\rm ln} (1-ux) {\rm ln} (1-ux) = 2\sum_{m=1}^\infty \sum_{n=1}^\infty \frac{1}{m(m+n)^3} = 2 \zeta (3,1),\ee
where the multiple zeta value (MZV) of depth 2 is defined by
\be \label{mzv} \zeta (a,b) = \sum_{m,n=1;m>n}^\infty \frac{1}{m^a n^b},\ee
which satisfies
\be \label{zetan1}
\zeta (n,1) = \frac{n}{2} \zeta (n+1) - \frac{1}{2} \sum_{k=1}^{n-2} \zeta (n-k) \zeta (k+1).\ee
Thus the integral \C{demo1} becomes $\zeta(4)/2$~\footnote{Note that for doing \C{demo1}, one does not have to introduce \C{mzv}. One can simply write 
\be 2\sum_{m=1}^\infty \sum_{n=1}^\infty \frac{1}{m(m+n)^3} = -2\zeta (4) + \sum_{m=1}^\infty \frac{1}{m} \zeta_H (3,m),\ee 
where $\zeta_H (3,m)$ is the Hurwitz zeta function. Then $\zeta (4)/2$ follows on using the
recusion relation
\be \zeta_H (3,m) = \zeta_H (3,m+1) + \frac{1}{m^3},\ee
and the summation relation
\be \sum_{m=1}^\infty \frac{1}{m} \zeta_H (3,m+1) = \frac{3}{2} \zeta (4) - \frac{1}{2} \zeta(2)^2 = \frac{1}{4} \zeta (4).\ee
However, MZV becomes very convenient for higher point amplitudes.
}.

The $\alpha'$ expansion for the 5 gluon amplitude can be carried out further. They lead to integrals like
\bea \label{d4f5} \int_0^1 dx \int_0^1 du \frac{1}{xu} {\rm ln} (1-ux) {\rm ln} u {\rm ln} x &=& -\zeta (5), \non \\ \int _0^1 dx \int_0^1 du \frac{1}{1-ux} ({\rm ln}u)^2 {\rm ln} (1-x) &=& 3 \zeta (2) \zeta (3)-6 \zeta(5), \eea
while at the next order they give integrals like
\bea \label{d6f5} \int_0^1 dx \int_0^1 du \frac{1}{ux} {\rm ln} (1-ux) ({\rm ln} u)^2 {\rm ln} x &=& 2\zeta (6), \non \\ \int_0^1 dx \int_0^1 du \frac{1}{ux} {\rm ln} (1-ux) {\rm ln} (1-x) ({\rm ln} u)^2 &=& \frac{7}{2} \zeta (6)- \zeta(3)^2,\eea
and integrals like
\bea \label{d8f5} \int_0^1 dx \int_0^1 du \frac{1}{xu} {\rm ln} (1-ux) ({\rm ln} u)^4 &=& -24 \zeta (7), \non \\ \int_0^1 dx \int_0^1 du \frac{1}{xu} {\rm ln} (1-ux) ({\rm ln}u)^3 {\rm ln} (1-x) &=& 6 \zeta (3) \zeta (4) + 6 \zeta (2) \zeta (5) - 24 \zeta (7)\non \\\eea
at the next higher order. These integrals are needed to calculate $L_1^{(1)}$ that follow from \C{intl} at higher orders in the $\alpha'$ expansion.

\subsection{Integrals for the six and seven gluon amplitude}

For the higher point multi--gluon amplitudes, there are a larger number of integrals to do at every order in the $\alpha'$ expansion. For our purposes, we mention only a few integrals involving the 6 and 7 gluon amplitudes, which are needed to derive the results in the main text. The other integrals yield similar answers. As before, all these integrals have a fixed transcendentality which is determined by the order of the $\alpha'$ expansion, or equivalently by the number of logarithmic terms in the integrand.

For the 6 gluon amplitude, some of the integrals are
\bea \int_0^1 dp \int_0^1 dy \int_0^1 dw \frac{1}{pwy} {\rm ln}(1-pwy) &=& -\zeta (4), \non \\ \int_0^1 dp \int_0^1 dy \int_0^1 dw \frac{1}{p(1-w)y} {\rm ln}(1-pwy) &=& \frac{5}{4} \zeta (4), \non \\ 
%\int_0^1 dp \int_0^1 dy \int_0^1 dw \frac{1}{pwy(1-py)} {\rm ln} (1-wy) &=& -\frac{5}{4} %\zeta (4), 
\int_0^1 dp \int_0^1 dy \int_0^1 dw  \frac{1}{pwy} {\rm ln} (1-pwy) {\rm ln} (1-p) &=& - \zeta (2)\zeta (3) + 3 \zeta (5), \non \\ \int_0^1 dp \int_0^1 dy \int_0^1 dw \frac{1}{pwy} {\rm ln} (1-w) {\rm ln} (1-py) &=& \zeta (2) \zeta (3) , \non \\ \int_0^1 dp \int_0^1 dy \int_0^1 dw \frac{1}{pwy}{\rm ln} (1-pwy) ({\rm ln} p)^2 &=& - 2\zeta (6) , \non \\ \int_0^1 dp \int_0^1 dy \int_0^1 dw \frac{1}{pwy} {\rm ln} (1-w) {\rm ln} (1-py) {\rm ln} (wy) &=& - \frac{7}{4} \zeta (6) - \zeta (3)^2, \non \\ \int_0^1 dp \int_0^1 dy \int_0^1 dw \frac{1}{pwy} {\rm ln} (1-pwy) ({\rm ln} p)^3 &=& 6 \zeta (7), \non \\ \int_0^1 dp \int_0^1 dy \int_0^1 dw \frac{1}{pwy} {\rm ln} (1-w) {\rm ln} (1-py) ({\rm ln} (wy))^2 &=& 4 \zeta (3) \zeta (4) + 2 \zeta (2) \zeta (5).\non \\ \eea

For the 7 gluon amplitude, some of the integrals are
\bea  \label{7pi}
\int_0^1 dp \int_0^1 dy \int_0^1 dw \int_0^1 dx \frac{1}{pwyx} {\rm ln}(1-pwyx) &=& - \zeta (5), \non \\  
%\int_0^1 dp \int_0^1 dy \int_0^1 dw \int_0^1 dx \frac{1}{py(1-wx)} {\rm ln} (1-pwy) &=& \ze%ta (2) \zeta (3) - 3\zeta (5), \non \\ 
\int_0^1 dp \int_0^1 dy \int_0^1 dw \int_0^1 dx \frac{1}{pw(1-xy)} {\rm ln} (1-pwx) {\rm ln} y&=& \frac{1}{2} \zeta(3)^2, 
%\non \\ \int_0^1 dp \int_0^1 dy \int_0^1 dw \int_0^1 dx \frac{1}{pwxy} {\rm ln} (1-py) {%\rm ln} (1-wxy) &=& \frac{1}{2} \zeta(3)^2, 
\non \\ \int_0^1 dp \int_0^1 dy \int_0^1 dw \int_0^1 dx \frac{1}{pwyx} {{\rm ln}p\rm ln} (1-ywx) {\rm ln} (1-py) &=& -4 \zeta (7) + 2 \zeta (2) \zeta (5). \non \\
\eea

\subsection{Integrals for the eight gluon amplitude}

For the 8 gluon amplitude, we list only a couple of integrals because in $L_1$ the 8 point amplitude integrals mostly yield terms which are subleading compared to the order in the $\alpha'$ expansion we are interested in. The integrals are  
\bea \label{8pi}
\int_0^1 dp \int_0^1 dy \int_0^1 dw \int_0^1 dx \int_0^1 dz \frac{1}{pwyxz} {\rm ln}(1-pwyxz) &=& - \zeta (6),\non \\ \int_0^1 dp \int_0^1 dy \int_0^1 dw \int_0^1 dx \int_0^1 dz \frac{1}{pywxz}  {\rm ln} (1-pyw) {\rm ln} (1-wzx)&=& 4 \zeta (7) - 2 \zeta (2) \zeta (5) . \non \\ \eea

As the number of gluons increase, the integrals get more and more complicated, and so we outline the details of one the integrals involved in the 7 gluon amplitude. The last integral in \C{7pi} is equal to\footnote{The last integral in \C{8pi} is the minus of this.} 
\be \label{genint}
-\int_0^1 \frac{dx}{x} Li_3^2 (x).\ee
Now
\bea \label{mzvint} \int_0^1 \frac{dx}{x} Li_3^2 (x) = \zeta (3) \zeta (4) - \zeta (2) \zeta (5) - \int_0^1 \frac{dx}{x} {\rm ln} (1-x) Li_5 (x),\eea
where we used the recurrence relation
\be x\frac{d}{dx} Li_n (x) = Li_{n-1} (x),\ee
integrated by parts, and substituted
\be Li_1 (x) = - {\rm ln} (1-x).\ee
The integral in \C{mzvint} is equal to
\be \sum_{m,n=1}^\infty \frac{1}{m^5 n(m+n)} = \zeta (5,2) +\zeta (2,5) + \zeta (4,3) +\zeta (3,4) + 2 \zeta (6,1) ,   \ee
on using \C{mzv}. Finally on using the MZV stuffle relation
\be \label{stuffle}\zeta (a,b) + \zeta (b,a) = \zeta (a) \zeta (b) - \zeta (a+b),\ee
and \C{zetan1}
we recover the answer in \C{7pi}.

Needless to say, as the number of gluons get larger, the integrals get more complicated, and performing the $\alpha'$ expansion gets more challenging. Generalities of doing these integrals in the context of string amplitudes have been discussed in~\cite{Schlotterer:2012ny,Drummond:2013vz,Broedel:2013tta,Broedel:2013aza}.

%\bibliographystyle{utphys}
%\bibliography{myrefs}
%\providecommand{\href}[2]{#2}\begingroup\raggedright\begin{thebibliography}{10}

\providecommand{\href}[2]{#2}\begingroup\raggedright\endgroup

%\end{thebibliography}\endgroup

\end{document}